\documentclass{article}

\usepackage[nonatbib,preprint]{neurips_2024}

\usepackage[utf8]{inputenc} 
\usepackage[T1]{fontenc}    
\usepackage{hyperref}       
\usepackage{url}            
\usepackage{booktabs}       
\usepackage{amsfonts}       
\usepackage{nicefrac}       
\usepackage{microtype}      
\usepackage{xcolor}         
\usepackage{algorithm}
\usepackage{algpseudocode}
\usepackage[numbers,sort&compress]{natbib}
\usepackage[labelfont=bf,textfont=it,font=small]{caption}

\usepackage{amsmath}
\usepackage{subcaption}
\usepackage{graphicx}
\usepackage{array}
\usepackage{enumitem}
\usepackage{xspace}
\usepackage{amssymb}  
\usepackage{amsfonts}  
\usepackage{pifont}    
\usepackage{titlesec}
\usepackage{multirow}
\usepackage{siunitx}
\usepackage{xspace}

\definecolor{darkgreen}{rgb}{0.0, 0.5, 0.0}

\titlespacing\section{0pt}{4pt plus 1pt minus 1pt}{3pt plus 1pt minus 1pt}
\titlespacing\subsection{0pt}{4pt plus 1pt minus 1pt}{3pt plus 1pt minus 1pt}
\titlespacing\subsubsection{0pt}{4pt plus 2pt minus 2pt}{3pt plus 2pt minus 2pt}

\hypersetup{colorlinks,
      linkcolor=blue,
      citecolor=blue,
      urlcolor=blue}

\newcommand{\pd}[2]{\frac{\partial #1}{\partial #2}} 
\newcommand{\tGD}{\tilde{\Gamma}_{D,h}}
\newcommand{\cref}[2]{\hyperref[#2]{#1~\ref*{#2}}}
\newcommand{\colref}[2]{\hyperref[#2]{#1~\ref*{#2}}}

\newcommand{\figref}[1]{\colref{Figure}{#1}}

\newcommand{\secref}[1]{\colref{Section}{#1}}
\newcommand{\tabref}[1]{\colref{Table}{#1}}
\newcommand{\coloredref}[2]{\hyperref[#2]{#1~\ref*{#2}}}
\newcommand{\coloredsubref}[3]{\hyperref[#2]{#1~\ref*{#2}{#3}}}
\newcommand{\Algref}[1]{\hyperref[#1]{Algorithm~\ref*{#1}}}
\newcolumntype{P}[1]{>{\centering\arraybackslash}p{#1}}

\newcommand{\Frontera}{\href{https://www.tacc.utexas.edu/systems/frontera}{Frontera}\xspace}

\newcommand{\FlowBench}{%
  \href{https://baskargroup.bitbucket.io/}{\textcolor{green!50!black}{FlowBench}}%
}

\title{\FlowBench{}: A Large Scale Benchmark for Flow Simulation over Complex Geometries}

\author{
  Ronak Tali\textsuperscript{1}\thanks{These authors contributed equally to this work.},
  Ali Rabeh\textsuperscript{1\fnsymbol{footnote}},
  Cheng-Hau Yang\textsuperscript{1\fnsymbol{footnote}},
  Mehdi Shadkhah\textsuperscript{1},
  Samundra Karki\textsuperscript{1}
  \ANDNEW
  Abhisek Upadhyaya\textsuperscript{2}, 
  Suriya Dhakshinamoorthy\textsuperscript{1},
  Marjan Saadati\textsuperscript{1},
  Soumik Sarkar\textsuperscript{1},
  \ANDNEW
  Adarsh Krishnamurthy\textsuperscript{1},
  Chinmay Hegde\textsuperscript{2},
  Aditya Balu\textsuperscript{1},
  Baskar Ganapathysubramanian\textsuperscript{1}
  \And
  \textsuperscript{1}Iowa State University\\
  \texttt{\{rtali,arabeh,chenghau,mehdish,}\\
  \texttt{samundra,snarayan,marjansd,soumiks,}\\
  \texttt{adarsh,baditya,baskarg\}@iastate.edu}\\
  \And
  \textsuperscript{2}New York University\\
  \texttt{\{au2216,chinmay.h\}@nyu.edu} 
}

\begin{document}

\maketitle


\begin{figure}[htbp]
    \centering
    \includegraphics[width=0.9\linewidth]{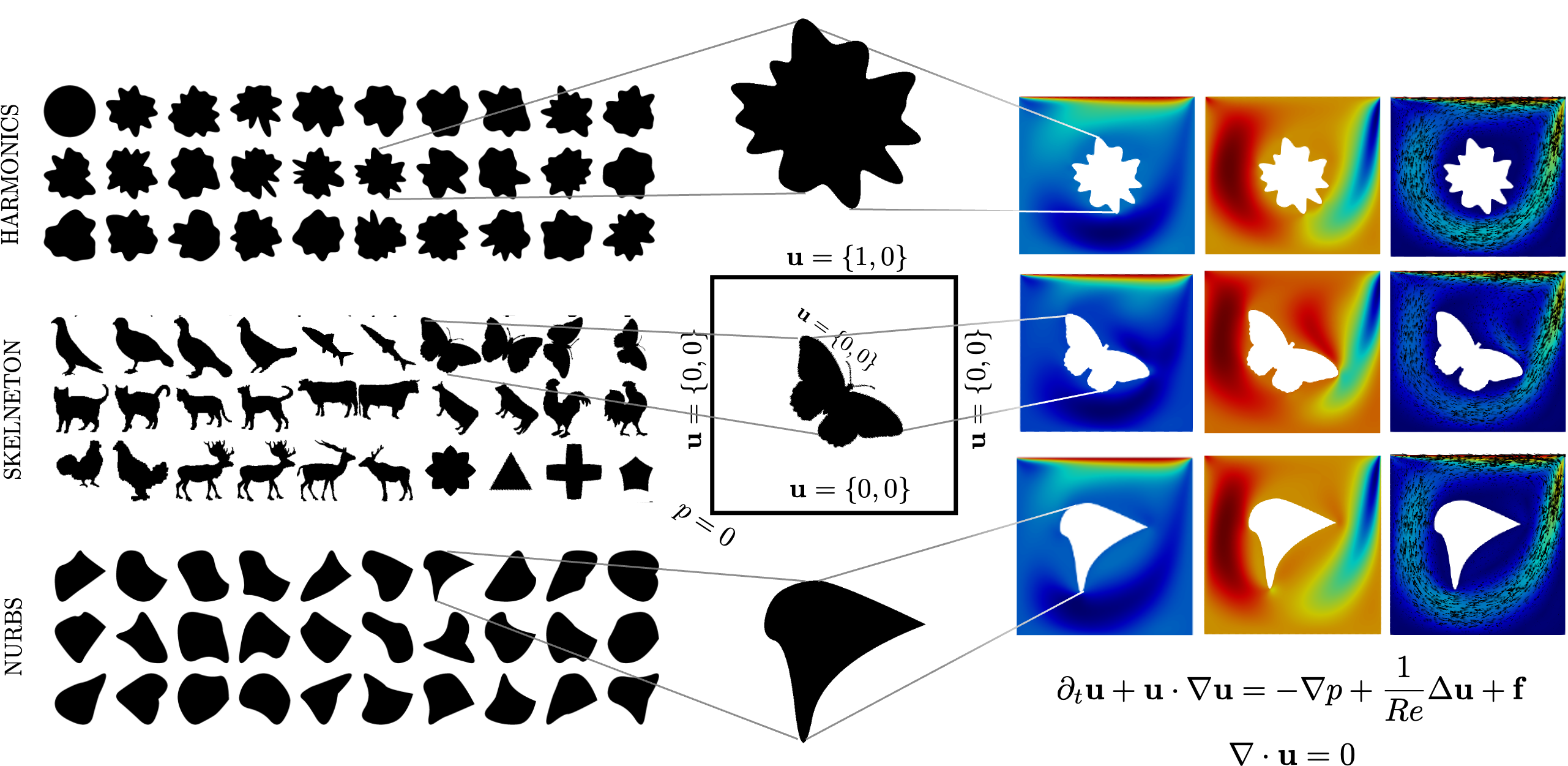}
    \caption{\FlowBench{} offers comprehensive datasets and metrics for assessing neural PDE solvers designed to model flow phenomena around complex objects. It includes three sets of application-relevant geometries with varying complexities and high-fidelity flow simulation data under different forcing conditions. The left panel in the figure above showcases 30 randomly selected shapes from each geometry group. The middle panel provides a close-up of one geometry within the computational domain, highlighting the boundary conditions. The right panel displays the simulation outputs, including velocity results for three samples.}
    \label{fig:poster}
\end{figure}

\begin{abstract}
Simulating fluid flow around arbitrary shapes is key to solving various engineering problems. However, simulating flow physics across complex geometries remains numerically challenging and computationally resource-intensive, particularly when using conventional PDE solvers. Machine learning methods offer attractive opportunities to create fast and adaptable PDE solvers. However, benchmark datasets to measure the performance of such methods are scarce, especially for flow physics across complex geometries. We introduce \FlowBench{}, a dataset for neural simulators with over 10K samples, which is currently larger than any publicly available flow physics dataset. \FlowBench{} contains flow simulation data across complex geometries (\textit{parametric vs. non-parametric}), spanning a range of flow conditions (\textit{Reynolds number and Grashoff number}), capturing a diverse array of flow phenomena (\textit{steady vs. transient; forced vs. free convection}), and for both 2D and 3D. \FlowBench{} contains over 10K data samples, with each sample the outcome of a fully resolved, direct numerical simulation using a well-validated simulator framework designed for modeling transport phenomena in complex geometries. For each sample, we include velocity, pressure, and temperature field data at 3 different resolutions and several summary statistics features of engineering relevance (such as coefficients of lift and drag, and Nusselt numbers). 
We envision that \FlowBench{} will enable evaluating the interplay between complex geometry, coupled flow phenomena, and data sufficiency on the performance of current, and future, neural PDE solvers. We enumerate several evaluation metrics to help rank order the performance of current (and future) neural PDE solvers. We benchmark the performance of several methods, including Fourier Neural Operators (FNO), Convolutional Neural Operators (CNO), DeepONets, and recent foundational models. This dataset (\href{https://huggingface.co/datasets/BGLab/FlowBench/tree/main}{here}) will be a valuable resource for evaluating neural PDE solvers that model complex fluid dynamics around 2D and 3D objects. 
\end{abstract}


\section{Introduction}

Accurate modeling of fluid flow around complicated objects is central to a plethora of applications. In aerospace and automotive applications, flow around wings, and car bodies can significantly impact design and performance~\cite {greenblatt2000control,spohn2002flow,you2008active}. In civil and environmental engineering applications, understanding fluid flow patterns around structures like buildings~\citep{ramponi2012cfd} and bridges~\citep{helgedagsrud2019ale} is critical from a safety perspective, while flow patterns inside buildings are important from a safety and comfort perspective~\cite{tan2023computational}. Bio-flow applications, such as flow around fish~\citep{seo2022improved,zhu2021numerical}, birds~\citep{song2014three}, insects~\citep{engels2016bumblebee}, and heart valves~\citep{xu2021computational}, are all important fields of study. Sports engineering scientists use fluid flow simulations to optimize equipment design, such as bike helmets and golf ball shapes~\citep{ting2003effects}. \textit{All these applications involve flow through and across \textbf{complex geometries}}.

Fluid flow is also often connected with thermal effects. A thermal mismatch between the fluid medium and the object geometry can produce buoyancy-driven phenomena, cause thermal plumes, and impact mixing and thermal transport with significant engineering implications, necessitating a multiphysics approach, where fluid flow and heat transfer are modeled concurrently. For instance, in the semiconductor industry, the design of heat exchangers must account for how the geometry influences both fluid flow and thermal transport~\citep{bhutta2012cfd,abeykoon2020compact}. Similarly, addressing urban heat island effects requires a deep understanding of the interaction between thermal and flow dynamics in complex urban settings~\citep{priyadarsini2008microclimatic,allegrini2018simulations}. Additionally, ensuring indoor comfort and safety, particularly in mitigating the transmission of infectious diseases, depends on accurately modeling the interplay of airflow and temperature within built environments~\citep{bhattacharyya2020novel,tan2023computational}. \textit{These examples further underscore the necessity of considering multiphyics flow simulations through and across \textbf{complex geometries}}.


A central challenge in accurately simulating flow patterns in complex geometries is the high resource cost of traditional simulation approaches. High-fidelity flow (and thermal) simulations often require hours to days (or sometimes even months~\citep{saurabh2023scalable}) on high-performance computing (HPC) systems. 
Scientific machine learning (SciML) has emerged as a  promising path towards resolving this challenge. By combining training data with domain-specific information (e.g., physical constraints and smoothness assumptions), SciML approaches offer fast simulation, better  extrapolation capabilities, and lower data requirements. This includes impactful applications like weather prediction~\citep{Rasp_2020} and canonical flows in simple geometries~\citep{bonnet2022airfrans, luo2024cfdbench, xu2023megaflow2d, janny2023eagle, bonnet2022extensible}. 

Despite these successes however, there is a notable lack of datasets for flow (and flow-thermal) interactions with complicated geometries. While databases exist for flow passing through simple shapes such as cylinders~\citep{luo2024cfdbench,xu2023megaflow2d} and airfoils~\citep{bonnet2022airfrans,bonnet2022extensible}, there remains a significant gap in datasets involving more diverse and complex shapes. We note the availability of a few datasets that include other geometries, but these geometries are limited to drone shapes~\citep{janny2023eagle}. 

In addition to the need for more complex geometries in existing flow datasets, there is also a shortage of multiphysics datasets. By multiphysics, we mean applications that are modeled as a set of (tightly) coupled partial differential equations (PDE), with each PDE modeling a specific physical phenomenon -- for instance, Navier-Stokes that models flow phenomena and the advective heat equation that models thermal phenomena. This gap exists because coupling different types of PDEs is inherently challenging and computationally expensive. Solving multiphysics problems requires sophisticated numerical methods and substantial computational resources to accurately simulate each subproblem and capture the interactions between various physical phenomena. 

\FlowBench{} seeks to enable the ML community to build the next generation of SciML neural PDE solvers by filling these gaps -- complex geometries and multiphysics phenomena. In particular, \FlowBench{} offers:
\begin{itemize}[topsep=0pt, itemsep=0pt, left=0pt] 
\item \textbf{Flow across complex geometries}: We simulate flow and thermal-flow phenomena across a wide range of complex shapes -- both parametric and non-parametric -- in 2D and 3D. These include simple shapes like ellipses, more complex blobs, and geometries like insects, animals, and birds. Flow across this spectrum of complex objects exhibits a rich array of vortex formation, flow separation, and a range of lift and drag profiles. The diverse shapes and the flow interactions with them provide a rich and intricate dataset for training geometry-aware SciML solvers that can be used for various applications.  


\item \textbf{Multiphysics simulations}: For each shape, we perform a variety of simulations representing flow (i.e., incompressible Navier-Stokes, NS) as well as thermal flow (i.e., coupled Navier-Stokes and Heat Transfer using the Bousinessque coupling, NS-HT) scenarios. Here, we span steady-state and transient behaviors, offering a comprehensive dataset and benchmarks to test and evaluate solvers under various scenarios. 
\begin{itemize}
    \item For the steady-state case, we consider a variant of the canonical fluid dynamics problem of lid-driven cavity flow (LDC), which is an example of internal flow. The cavity has an object (complex geometry) placed inside (see \figref{fig:reynolds_drag_lift}). Additionally, we consider a temperature difference between the cavity walls and the object, producing thermal and flow coupling. For each object, we simulate across a range of Reynolds number $Re \in [10^1, 10^3]$, and Grashof number $Gr \in [10^1, 10^7]$. This range offers a variety of forced \(\left({{Gr}}/{{Re}^2} \approx 0.1\right)\), mixed, and free convection  \(\left({{Gr}}/{{Re}^2} \approx 10\right)\) scenarios. In this case, there are nearly 9000 unique samples for 2D and 500 unique samples for 3D.
    \item For the transient case, we consider a variant of another canonical fluid dynamics problem -- flow past a bluff object (FPO) -- which is an example of external flow. Here, we consider flow moving past a stationary (complex geometry) object that is placed in a large domain. The fluid exhibits various intricate time-dependent patterns as it moves past the object. For each object, we simulate across a range of Reynolds number $Re \in [10^2, 10^3]$, offering an array of vortex-shedding frequencies and other time-dependent patterns. In this case, there are over 1000 unique samples in 2D.
\end{itemize}
We provide field data of velocity, pressure, and temperature for each of the over $10K$ simulations. Additionally, we provide summary engineering features, including the coefficient of lift ($C_L$), drag ($C_D$), and the average heat transfer (Nusselt number, $Nu$) from the surface of the complex object. We provide our dataset on huggingface at \url{https://huggingface.co/datasets/BGLab/FlowBench/tree/main} as a benchmark for others interested in the development and evaluation of SciML models. 

\item \textbf{Benchmark metrics and comparisons}: Besides the dataset, we also include workflows to train select group of neural operators -- Fourier Neural Operators (FNO), Convolutional Neural Operators (CNO), and Deep Operator Networks (DeepONets). Neural operators can be trained using our code for the steady-state case and study the comprehensive evaluation metrics that we recommend in \secref{subsec:metrics}. We also suggest a hierarchy of in-distribution and out-of-distribution tests to evaluate the generalizability of these models. 
\end{itemize}

The dataset (consisting of over $10K$ samples), evaluation metrics, workflows, and trained models together make \FlowBench{} a valuable tool for the ML community to create SciML solvers of coupled phenomena involving complex geometries. All data, visualizations, models, and model evaluations are available on the \FlowBench~\href{https://baskargroup.bitbucket.io/}{website}.

\begin{table}[t!]
\centering
\caption{Comparison of incompressible flow simulation data in AirfRANS, Curated Dataset, CFDBench, MegaFlow, Eagle, Graph-Mesh, PDEBench, and our \FlowBench{}. An orange check mark indicates a restricted family of shapes (Airfoil: AirRANS and Graph-Mesh; Cylinder: CFDBench, Cylinders and Ellipses: MegaFlow; Drone geometry: Eagle).}
\label{tab:comparison}
\setlength\extrarowheight{2pt}
\begin{tabular}{|>{\centering}p{3.5cm}|>{\centering}p{2.0cm}|>{\centering}p{2.0cm}|>{\centering}p{2.0cm}|>{\centering\arraybackslash}p{2.0cm}|}
\hline
\textbf{Name} & \textbf{Dimensions} & \textbf{\# Simulations} & \textbf{Geometry}
& \textbf{Multiphysics}\\ \hline
\textbf{AirfRANS}~\citep{bonnet2022airfrans} &  2  & 1000  & \textcolor{orange!90!black}{\checkmark} & \textcolor{red}{\ding{55}}
\\
\textbf{Curated Dataset}~\citep{mcconkey2021curated} & 2  & 116 & \textcolor{red}{\ding{55}} & \textcolor{red}{\ding{55}}
\\
\textbf{CFDBench}~\citep{luo2024cfdbench} & 2  & 739 & \textcolor{orange!90!black}{\checkmark}  & \textcolor{red}{\ding{55}}
\\
\textbf{MegaFlow}~\citep{xu2023megaflow2d} & 2  & 3000  & \textcolor{orange!90!black}{\checkmark}  & \textcolor{red}{\ding{55}}
\\
\textbf{Eagle}~\citep{janny2023eagle} & 2  & 1200  & \textcolor{orange!90!black}{\checkmark}  & \textcolor{red}{\ding{55}}
\\
\textbf{Graph-Mesh}~\citep{bonnet2022extensible} & 2  & 230 & \textcolor{orange!90!black}{\checkmark}  & \textcolor{red}{\ding{55}}
\\
\textbf{PDEBench}~\citep{takamoto2023pdebench} & 2  & 1000  & \textcolor{red}{\ding{55}} & \textcolor{red}{\ding{55}}
\\
\textbf{ScalarFlow}~\citep{eckert2019scalarflow} & 3  & 100  & \textcolor{red}{\ding{55}} & \textcolor{darkgreen}{\checkmark}
\\
\textbf{Fluid Flow Dataset}~\citep{jakob2020fluid} & 2  &  8000  & \textcolor{red}{\ding{55}} & \textcolor{red}{\ding{55}}
\\
\textbf{BubbleML}~\citep{hassan2023bubbleml} & 2,3 & 79 & \textcolor{red}{\ding{55}}  & \textcolor{darkgreen}{\checkmark}
\\
\hline
\textbf{\FlowBench{} (ours)} & 2,3  & \textbf{10650}
 &  \textcolor{darkgreen}{\checkmark} & \textcolor{darkgreen}{\checkmark}
\\
\hline
\end{tabular}
\end{table}

\section{Related Work}
\label{related_work}

Most publicly available benchmark datasets are summarized and compared in \tabref{tab:comparison}.  PDEBench~\citep{takamoto2023pdebench} presents a dataset consisting of simulations run for a wide range of PDEs, not just flow physics. It includes simulations for both compressible and incompressible Navier-Stokes problems, in both two and three dimensions.

CFDBench~\citep{luo2024cfdbench} is a fluid flow-focused dataset containing a total of 739 cases. These cases are distributed across various setups: lid-driven cavity (with varying density and viscosity over 25 different length and width combinations), tube flow (varying density, viscosity, and geometry, with 50 different inlet velocity conditions), dam flow, and cylinder flow specialties.

Megaflow2D~\citep{xu2023megaflow2d} is a comprehensive collection of 3000 cases for 2D Navier-Stokes problems. Each case features different geometrical configurations, including circles, ellipses, and nozzles. All simulations are performed at a fixed Reynolds number of 300.

\citet{mcconkey2021curated} focuses exclusively on 2D turbulence simulations, covering 841 different cases across 29 flow scenarios such as a periodic hill, square duct, parametric bump, converging-diverging channel, and curved backward-facing step. Each scenario includes 29 simulations with varying parameters, such as Reynolds numbers, producing 841 data samples.

The Graph-Mesh~\citep{bonnet2022extensible} dataset simulates incompressible 2D Navier-Stokes problems at high Reynolds numbers (greater than $10^6$) exclusively for Aerofoils. It varies the angle of attack, inlet velocity, Reynolds number, and Mach number for a total of 230 simulations. A distinguishing feature is the calculation of drag and lift after fitting the SciML models, not just the solutions.

WeatherBench~\citep{Rasp_2020} condenses raw weather data from the ERA5 dataset by reducing resolution levels to fit within a GPU. Along with discrete weather-based variables, WeatherBench runs simulations for a range of $u,v$ velocities, temperature, and vorticity. It uses the entire world as its grid and provides both 2D and 3D data post-processed from the ERA5 dataset.

ScalarFlow~\citep{eckert2019scalarflow} is the multiphysics simulation dataset coupling Navier-Stokes with density distribution solving, focusing on buoyancy-driven smoke plume reconstructions. It provides volumetric 3D flow reconstructions for complex buoyancy-driven flows transitioning to turbulence.

The Fluid Flow Dataset~\citep{jakob2020fluid} contains 8000 unsteady 2D fluid flow simulations, each with 1001 time steps, parameterized by Reynolds number.

BubbleML~\citep{hassan2023bubbleml} is another multiphysics dataset to couple two-phase Navier-Stokes and energy equations. It includes 79 simulations covering nucleate pool boiling, flow boiling, and sub-cooled boiling under various gravity conditions, flow rates, sub-cooling levels, and wall superheat. 

EAGLE~\citep{janny2023eagle} is a large (over 1.1M 2D simulations) dataset that models the airflow generated by 2D UAV moving in a 2D environment with different boundary geometries. This dataset captures the highly turbulent flow consisting of non-periodic eddies induced by the varying geometries.

\section{FlowBench}



\subsection{Geometries}

\begin{figure}[t!]
    \begin{subfigure}{0.1\textwidth}
    \centering
    \raisebox{2.5\height}{\makebox[0.1\textwidth][c]{\textbf{G1}}} 
    \end{subfigure}%
    \begin{subfigure}{0.1\textwidth}
    \centering
        \includegraphics[width=\textwidth]{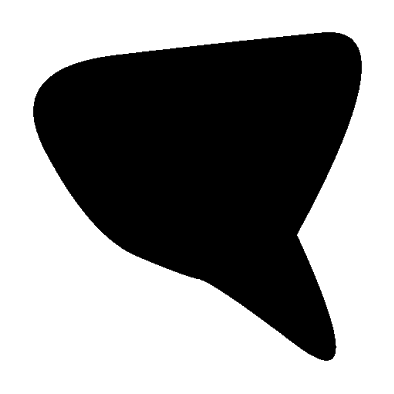}
    \end{subfigure}%
    \begin{subfigure}{0.1\textwidth}
    \centering
        \includegraphics[width=\textwidth]{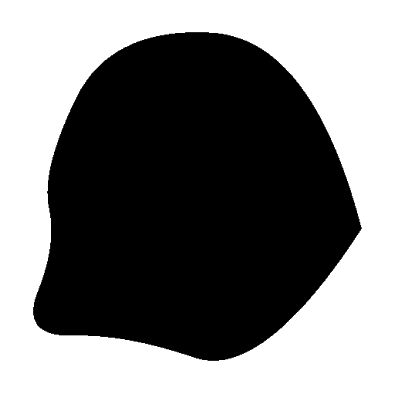}
    \end{subfigure}%
    \begin{subfigure}{0.1\textwidth}
    \centering
        \includegraphics[width=\textwidth]{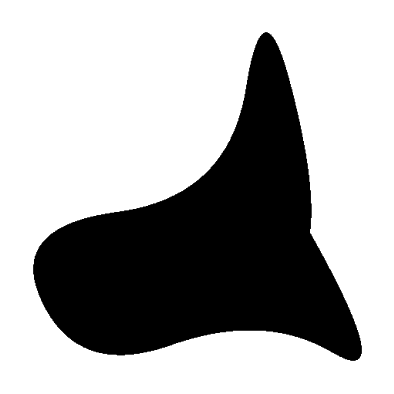}
    \end{subfigure}%
    \begin{subfigure}{0.1\textwidth}
    \centering
        \includegraphics[width=\textwidth]{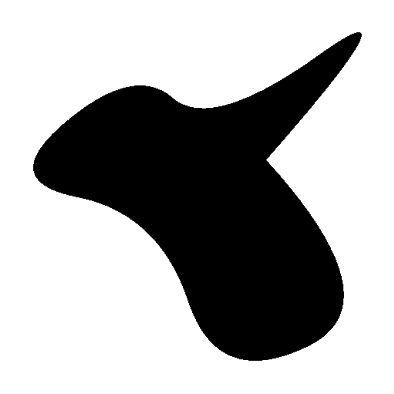}
    \end{subfigure}%
    \begin{subfigure}{0.1\textwidth}
    \centering
        \includegraphics[width=\textwidth]{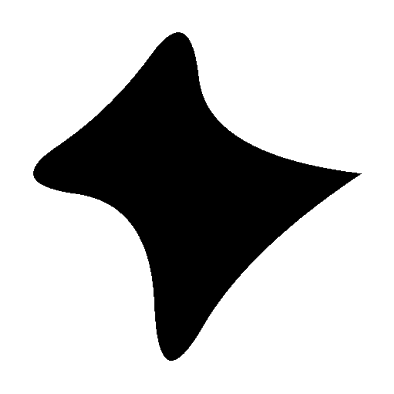}
    \end{subfigure}%
        \begin{subfigure}{0.1\textwidth}
    \centering
        \includegraphics[width=\textwidth]{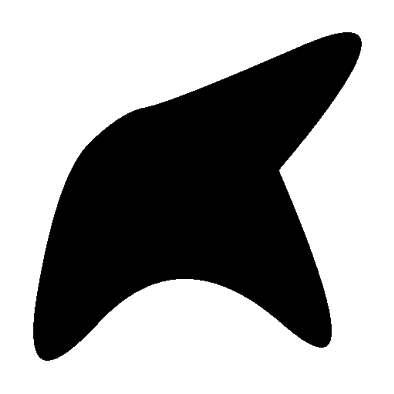}
    \end{subfigure}%
        \begin{subfigure}{0.1\textwidth}
    \centering
        \includegraphics[width=\textwidth]{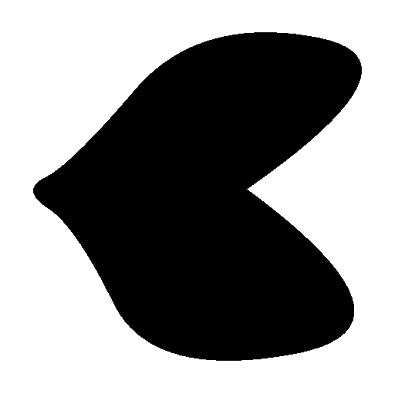}
    \end{subfigure}%
        \begin{subfigure}{0.1\textwidth}
    \centering
        \includegraphics[width=\textwidth]{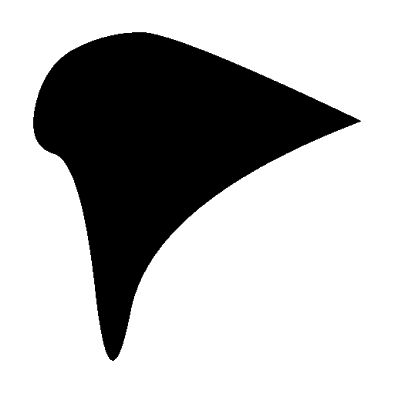}
    \end{subfigure}%
        \begin{subfigure}{0.1\textwidth}
    \centering
        \includegraphics[width=\textwidth]{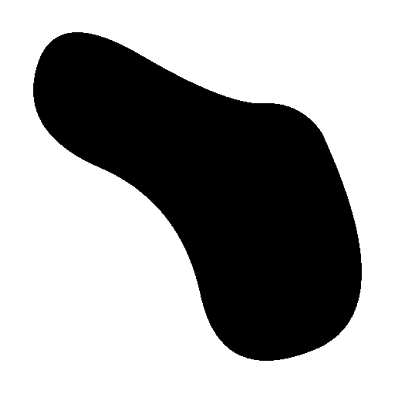}
     \end{subfigure}\\
    \begin{subfigure}{0.1\textwidth}
    \centering
    \raisebox{2.5\height}{\makebox[0.1\textwidth][c]{\textbf{G2}}} 
    \end{subfigure}%
    \begin{subfigure}{0.1\textwidth}
    \centering
        \includegraphics[width=\textwidth]{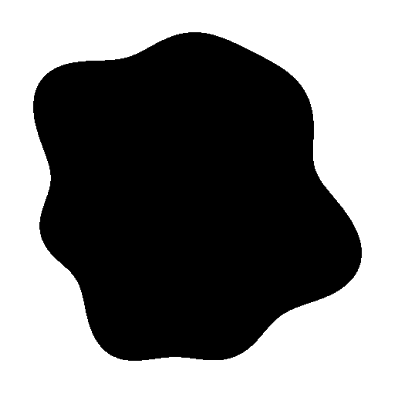}
    \end{subfigure}%
    \begin{subfigure}{0.1\textwidth}
    \centering
        \includegraphics[width=\textwidth]{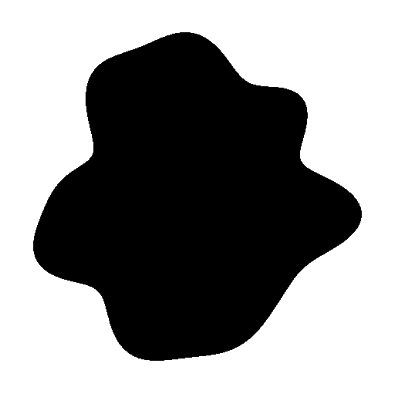}
    \end{subfigure}%
    \begin{subfigure}{0.1\textwidth}
    \centering
        \includegraphics[width=\textwidth]{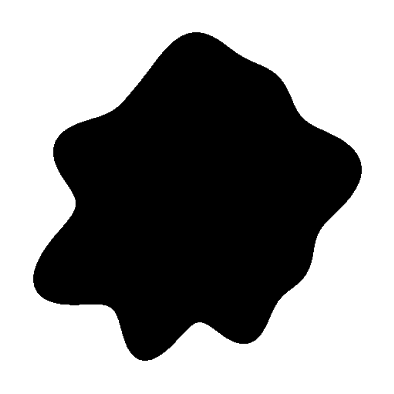}
    \end{subfigure}%
    \begin{subfigure}{0.1\textwidth}
    \centering
        \includegraphics[width=\textwidth]{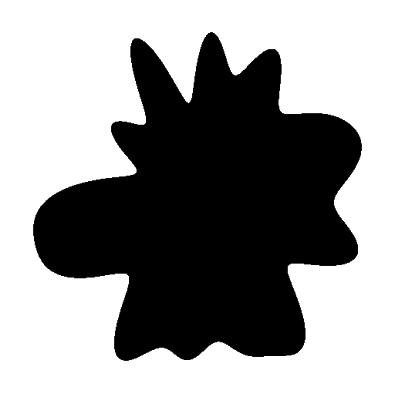}
    \end{subfigure}%
    \begin{subfigure}{0.1\textwidth}
    \centering
        \includegraphics[width=\textwidth]{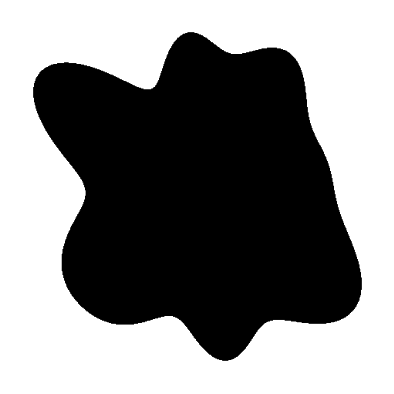}
    \end{subfigure}%
        \begin{subfigure}{0.1\textwidth}
    \centering
        \includegraphics[width=\textwidth]{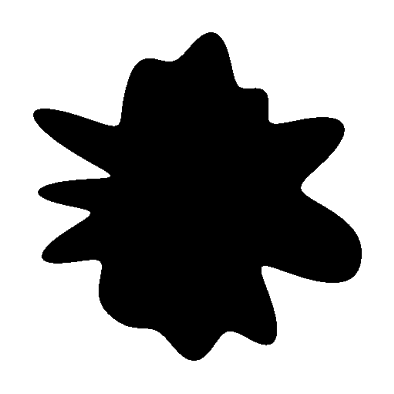}
    \end{subfigure}%
        \begin{subfigure}{0.1\textwidth}
    \centering
        \includegraphics[width=\textwidth]{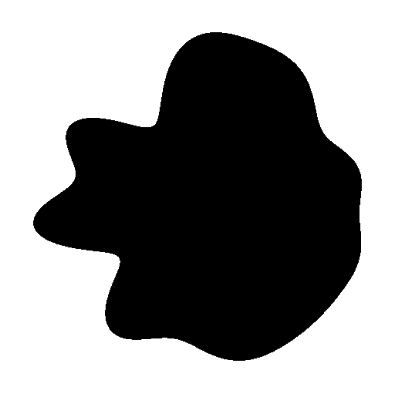}
    \end{subfigure}%
        \begin{subfigure}{0.1\textwidth}
    \centering
        \includegraphics[width=\textwidth]{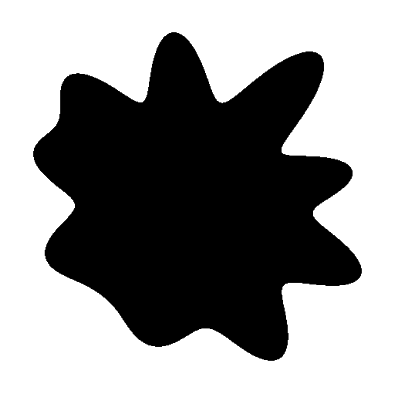}
    \end{subfigure}%
        \begin{subfigure}{0.1\textwidth}
    \centering
        \includegraphics[width=\textwidth]{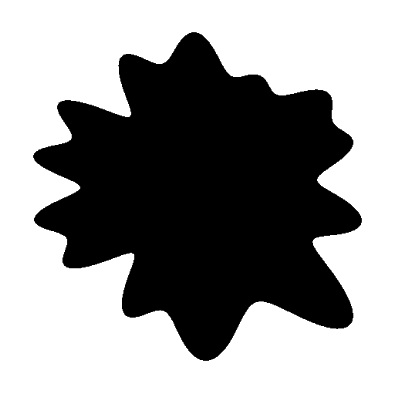}
    \end{subfigure}\\
    \begin{subfigure}{0.1\textwidth}
    \centering
    \raisebox{2.5\height}{\makebox[0.1\textwidth][c]{\textbf{G3}}} 
    \end{subfigure}%
    \begin{subfigure}{0.1\textwidth}
        \includegraphics[width=\textwidth]{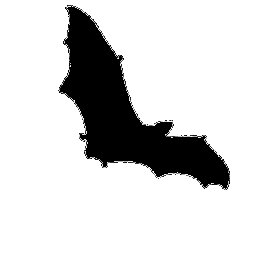}
    \end{subfigure}%
\begin{subfigure}{0.1\textwidth}
    \centering
        \includegraphics[width=\textwidth]{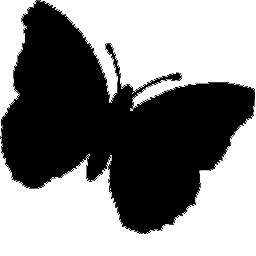}
    \end{subfigure}%
    \begin{subfigure}{0.1\textwidth}
    \centering
        \includegraphics[width=\textwidth]{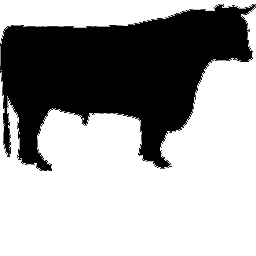}
    \end{subfigure}%
    \begin{subfigure}{0.1\textwidth}
    \centering
        \includegraphics[width=\textwidth]{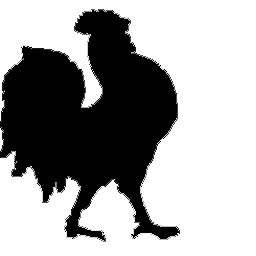}
    \end{subfigure}%
    \begin{subfigure}{0.1\textwidth}
    \centering
        \includegraphics[width=\textwidth]{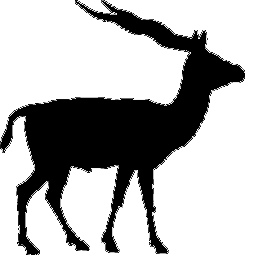}
    \end{subfigure}%
    \begin{subfigure}{0.1\textwidth}
    \centering
        \includegraphics[width=\textwidth]{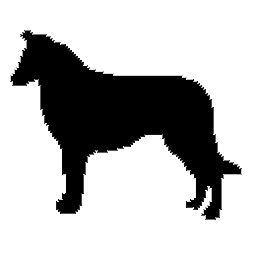}
    \end{subfigure}%
    \begin{subfigure}{0.1\textwidth}
    \centering
        \includegraphics[width=\textwidth]{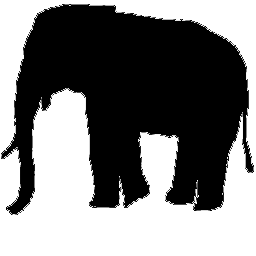}
    \end{subfigure}%
    \begin{subfigure}{0.1\textwidth}
    \centering
        \includegraphics[width=\textwidth]{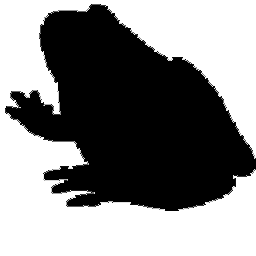}
    \end{subfigure}%
    \begin{subfigure}{0.1\textwidth}
    \centering
        \includegraphics[width=\textwidth]{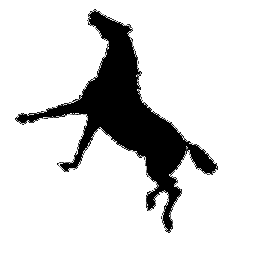}
    \end{subfigure}%
    \caption{Examples of the diverse and complex geometries in \FlowBench{} using 9 samples from each of the three groups. The first row corresponds to geometries from the nurbs group G1, the second row to the spherical harmonics group G2, and the third row to the skelneton group G3.}
    \label{fig:geometry}   
\end{figure}

Our dataset includes three distinct categories of geometries, namely \textbf{G1}, \textbf{G2}, and \textbf{G3} as illustrated in \figref{fig:geometry}. The first set of geometries, \textbf{G1}, consists of parametric shapes generated using Non-Uniform Rational B-Splines (NURBS) curves. NURBS are mathematical representations used in computer graphics and CAD systems to generate and represent curves and surfaces. They offer great flexibility and precision in modeling complex shapes. Each NURBS curve is defined by a set of control points, the degree of the basis function, and knot vectors \cite{piegl2012nurbs}. We use a uniform knot vector with a second-order (quadratic) basis function, which remains fixed. However, the positions of eight control points are randomly varied to produce a variety of curves. We ensure that the shapes are smooth and do not have any discontinuities or self-intersections. All shapes are normalized to fit within the unit hypercube, $[0, 1]^2$. We provide the shapes as well as code for recreating these geometries. 

The next set of geometries, \textbf{G2}, consists of parametric shapes generated using spherical harmonics~\citep{Wei2018}. We randomly select $N = 8,..., 15$ harmonics with amplitudes ($a_n$, $b_n$) ranging from 0 to 0.2. The radial function $r(t) = 0.5 + \sum_{n=1}^{N} \left( a_n \cos(nt) + b_n \sin(nt) \right)$ then defines the shape; and is computed at 500 evenly spaced points in $t \in [0, 2\pi]$. We normalize \(r(t)\) so that any surface point is within a distance of 0.5 from the center of the shape, $r(t) = 0.5 \left( \frac{r(t)}{r_{\text{max}}} \right)$. We provide shapes as well as code for recreating these geometries.

The last set of geometries, \textbf{G3}, consist of non-parametric shapes sampled from the grayscale dataset in SkelNetOn~\citep{demir2019skelneton,Atienza2019}. We apply a Gaussian blur filter with a scale of 2 to smoothen out some of the thin features of the object. This ensures the shape remains consistent across the three resolutions we provide data for. We then scale the shape to ensure it is contained in the unit hypercube $[0, 1]^2$.

\subsection{Flow Physics Problem: Domain,  Boundary Conditions, and Outputs}\label{subsec:composition}

\begin{figure}[b!]
    \centering
    \includegraphics[width=0.95\linewidth,trim=0 0 0.5in 0,clip]{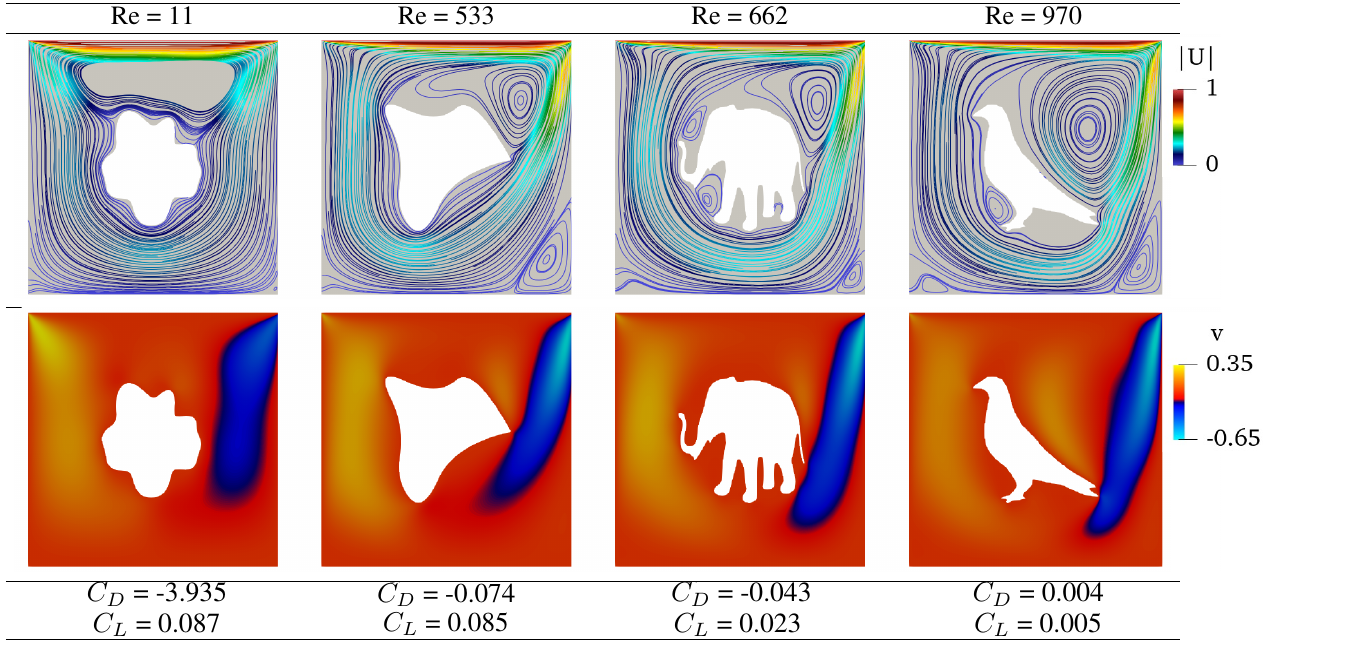} 
    \caption{Drag coefficients ($C_D$), and lift coefficients ($C_L$) for different shapes and different Reynolds numbers in pure-NS LDC simulations.}
    \label{fig:reynolds_drag_lift}
\end{figure}

In the 2D LDC setup, a square features three stationary walls and one moving lid, with a domain size of [0, 2] × [0, 2]. An object is placed in the middle of the flow within the chamber. Examples of LDC simulations showing streamlines and y-direction velocities, along with the drag coefficient ($C_D$) and lift coefficient ($C_L$) values, are shown in \figref{fig:reynolds_drag_lift}. Increasing the Reynolds number brings the vortices closer to the right wall, with additional vortices forming at the bottom-left and bottom-right corners. With increasing Reynolds numbers, the smaller viscous forces acting on the geometries decrease both $C_D$ and $C_L$.

\begin{figure}[t!]
    \centering
    \includegraphics[width=0.95\linewidth,trim=0 0 0 0,clip]{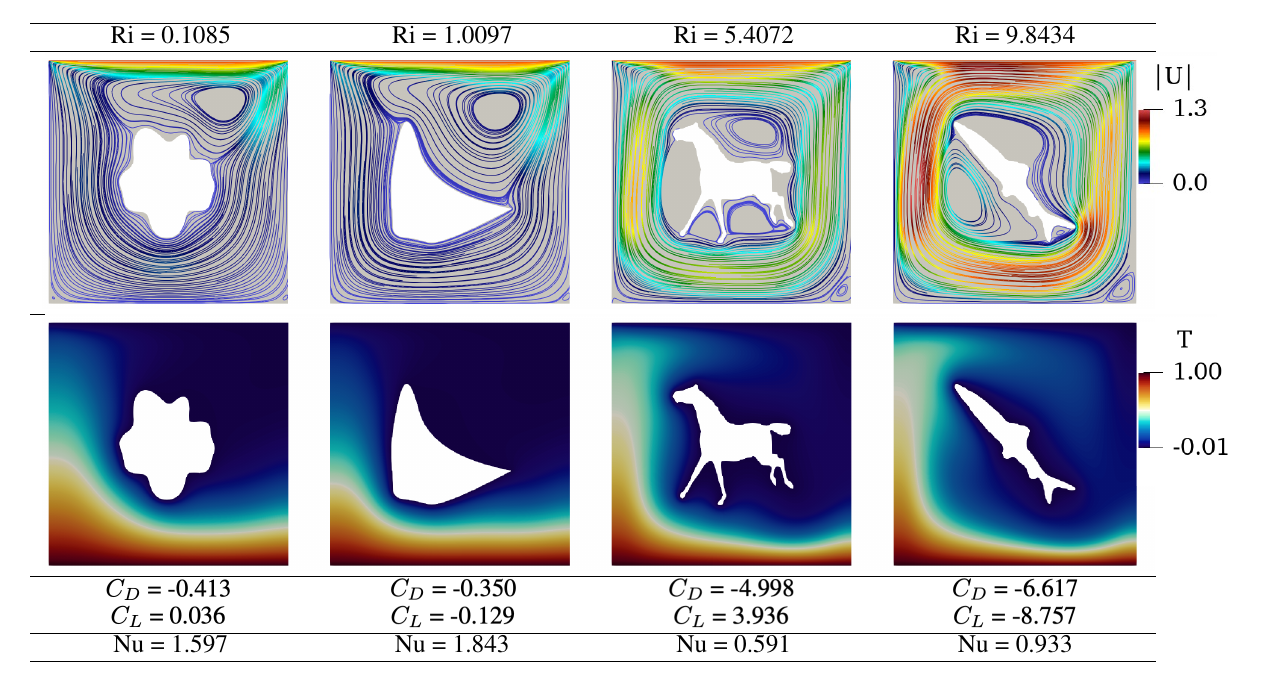} 
    \caption{Drag coefficients ($C_D$), lift coefficients ($C_L$), and Nusselt numbers (Nu) for different shapes and different Richardson numbers in NSHT LDC simulations (fixed Re = 100).}
    \label{fig:richardson_drag_lift_nusselt}
\end{figure}

For the 2D Navier-Stokes and heat transfer (NSHT) lid-driven cavity (LDC) problem, the bottom wall is set to a temperature of 1, while the top wall is set to 0. The left and right walls have zero-flux temperature boundary conditions. The object surface is set to a temperature of 0. This setup produces a combination of forced convection (flow due to the moving lid) and Rayleigh-Benard instabilities (flow due to buoyancy-driven natural convection).
Examples of NSHT-LDC simulations with fixed Reynolds number, showing streamlines, temperatures, and the values of $C_D$, $C_L$, and Nu, are presented in \figref{fig:richardson_drag_lift_nusselt}. We use a non-dimensional number, Richardson number, defined as $Ri = Gr/Re^2$, representing the ratio between buoyancy and inertial force. As $Ri$ increases, we observe higher values of \(C_D\) and \(C_L\). A higher Richardson number indicates that buoyancy effects are more significant than the forced flow. The heated fluid rises more strongly, creating greater circulation within the chamber. The increased circulation results in stronger forces acting on the object, leading to higher \(C_D\) and \(C_L\).

\begin{figure}[t!]
    \centering
    \includegraphics[width=0.95\linewidth,trim=0 0 0 0,clip]{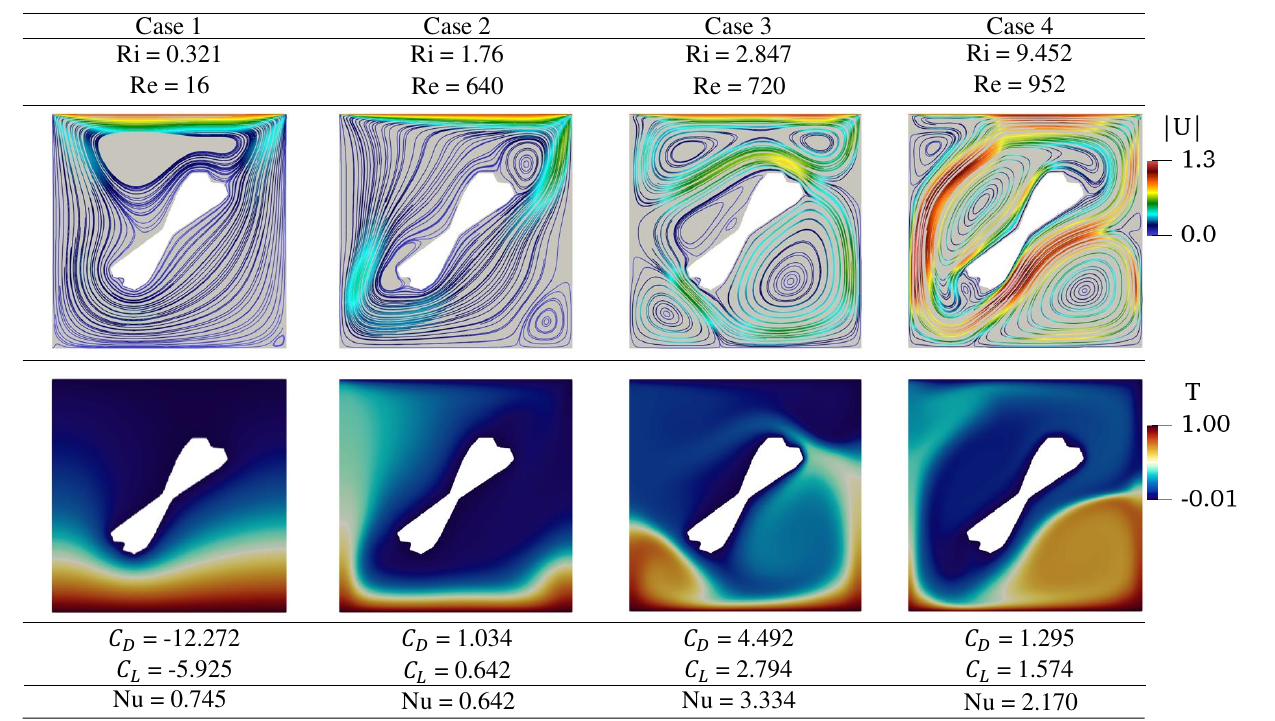} 
    \caption{Drag coefficients ($C_D$), lift coefficients ($C_L$), and Nusselt numbers (Nu) for different shapes and different Richardson numbers in NSHT LDC simulations (random Reynolds number).}
    \label{fig:richardson_drag_lift_nusselt_randomRe}
\end{figure}

Apart from the NSHT-LDC simulations with a fixed Reynolds number, we also conduct simulations with randomly varying Reynolds and Richardson numbers. We pick up one geometry to demonstrate in \figref{fig:richardson_drag_lift_nusselt_randomRe}. This showcase examines four cases with progressively increasing Reynolds and Richardson numbers. In Case 1, the high-speed flow predominantly occurs at the top of the geometry, resulting in a negative \(C_D\) due to the flow applying force on the upper part of the geometry. As we increase the Reynolds number in Case 2, the flow shifts towards the bottom part of the geometry, applying force in the positive y-direction, leading to a change in the sign of \(C_D\) from Case 1 to Case 2. The same as pure-NS flow, we observe a high-speed jet in the y-direction near the left-hand side wall due to the higher Reynolds number in Case 2. The magnitudes of \(C_D\) and \(C_L\) decrease from Case 1 to Case 2 because the viscous forces are reduced as the Reynolds number increases. Given that the Reynolds numbers in Case 2 and Case 3 are close, the reduction in force due to the increasing Reynolds number is mitigated, and the Richardson number plays a crucial role in enhancing flow circulation, leading to an increase in \(C_D\) and \(C_L\). However, from Case 3 to Case 4, the drag force decreases due to the higher Reynolds number, as observed in \figref{fig:reynolds_drag_lift}. Additionally, as the Richardson number increases, the temperature convect to higher positions, progressing from Case 1 to Case 4.

\begin{figure}[b!]
    \centering
    \begin{subfigure}[b]{0.4\linewidth}
        \includegraphics[width=\linewidth,trim=0 0 0 0,clip]{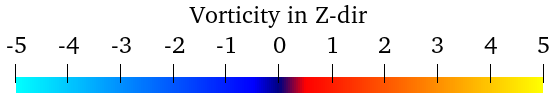}
    \end{subfigure}\\
    \begin{subfigure}{0.45\textwidth}
        \includegraphics[width=\linewidth,trim=0 280 0 280,clip]{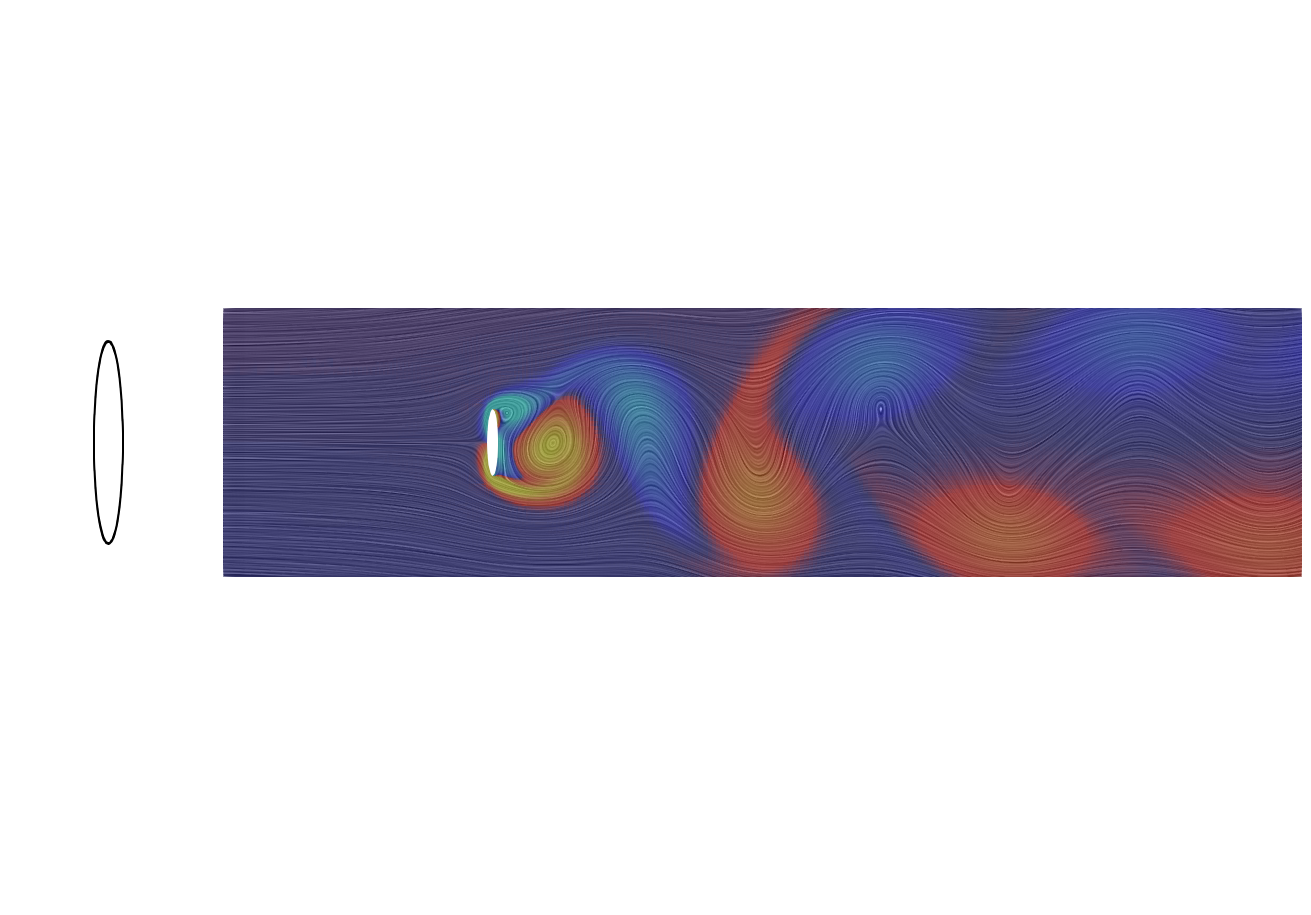}
        \caption{Re = 100}
        \label{fig:Re_100}
    \end{subfigure}
    \hfill
    \begin{subfigure}{0.45\textwidth}
    \centering
        \includegraphics[width=\linewidth,trim=0 280 0 280,clip]{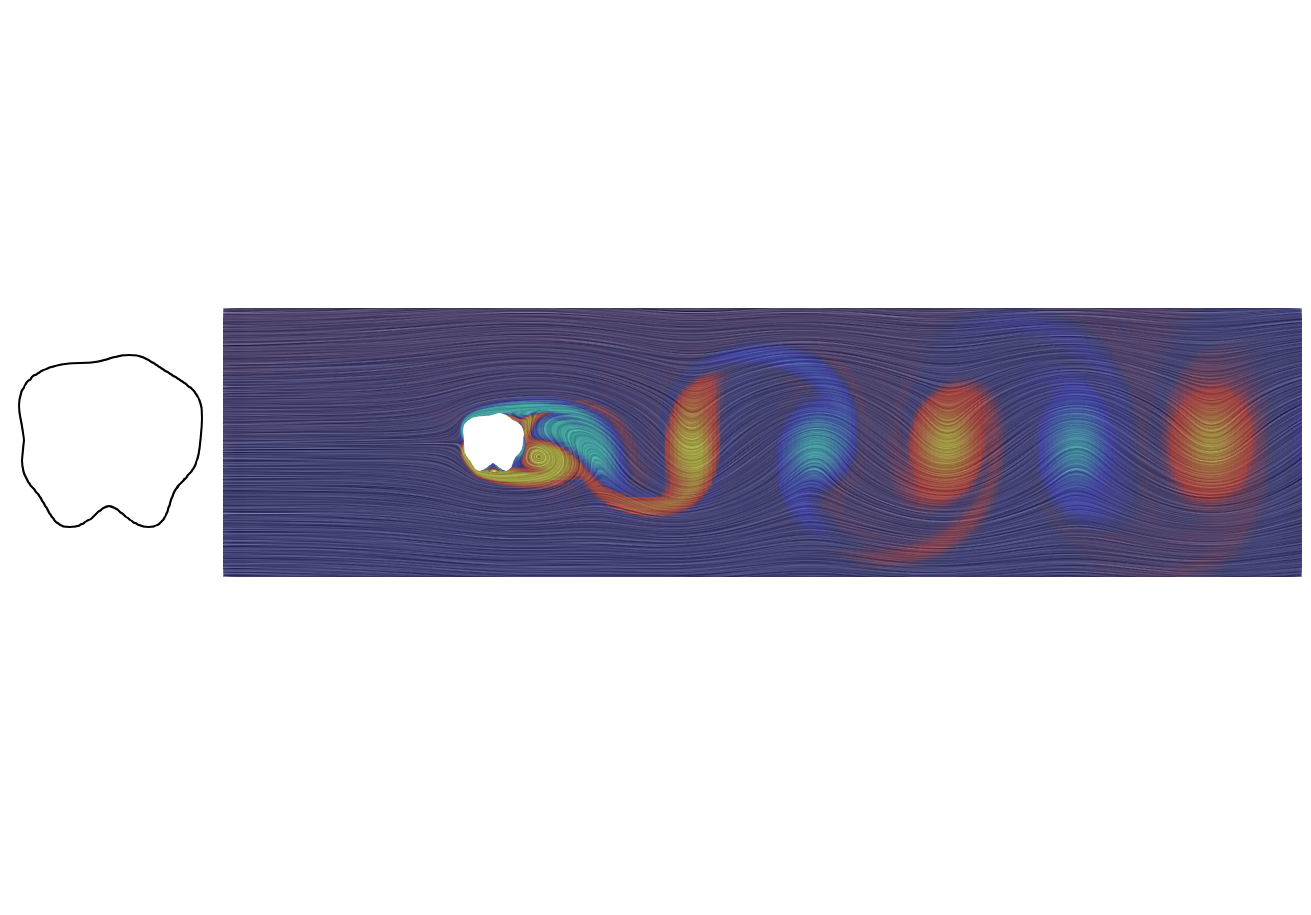}
        \caption{Re = 534}
        \label{fig:Re_534}
    \end{subfigure}%
    \\
    \begin{subfigure}{0.45\textwidth}
    \centering
        \includegraphics[width=\linewidth,trim=0 280 0 280,clip]{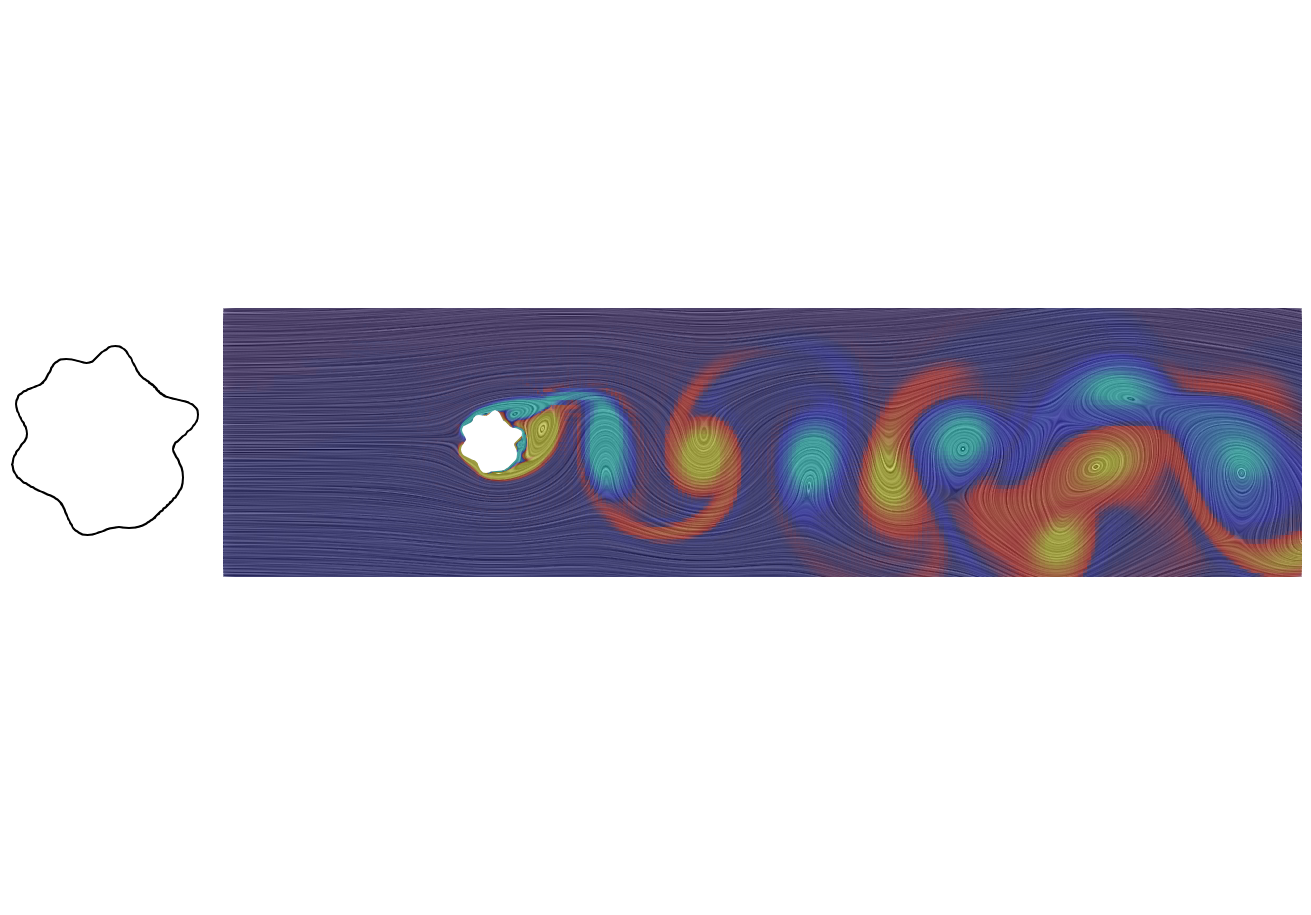}
        \caption{Re = 700}
        \label{fig:Re_700}
    \end{subfigure}
    \hfill
    \begin{subfigure}{0.45\textwidth}
    \centering
        \includegraphics[width=\linewidth,trim=0 280 0 280,clip]{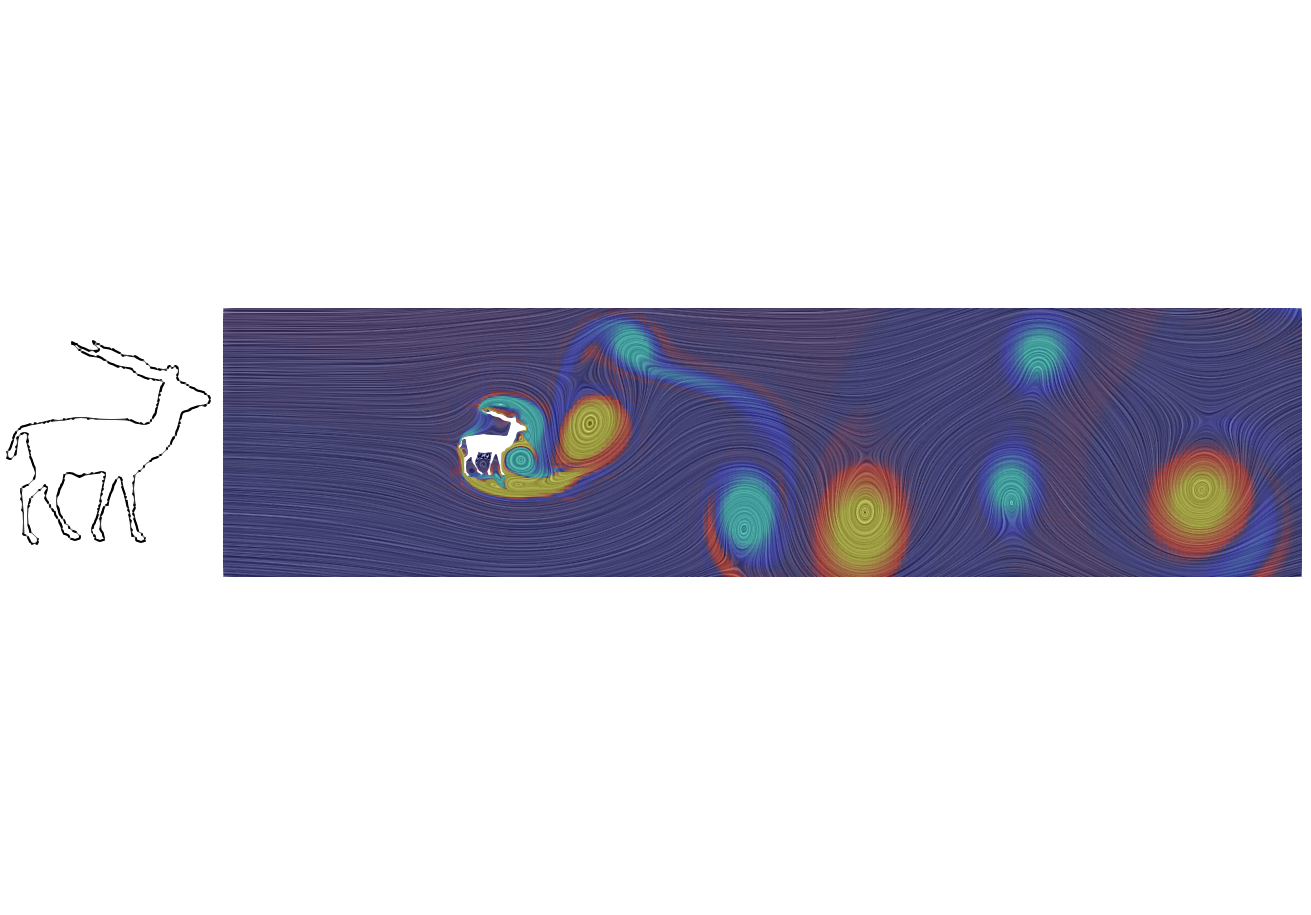}
        \caption{Re = 1000}
        \label{fig:Re_1000}
    \end{subfigure}%
    \caption{Time snapshots of vorticity for the FPO problem, illustrating vortex shedding profiles across four samples of geometries and Reynolds numbers. Each sample showcases distinct flow characteristics and vortex dynamics.}
    \label{fig:FlowCharacteristics_FPO_NS}
\end{figure}

For the FPO setup, we consider a domain $[0, 64] \times [0, 16]$ with the object placed at (6, 8). A parabolic velocity inlet boundary condition is applied on the left, no-slip boundary conditions are applied on the top and bottom walls, and a zero pressure boundary condition is applied on the right. The large domain size for the FPO problem is chosen to capture as much physical detail as possible in our dataset. We report a smaller cropped-out region of size $[0, 16] \times [0, 4]$ from this dataset, representing a tradeoff between dataset size and the amount of physics captured. \figref{fig:FlowCharacteristics_FPO_NS} shows time snapshots of representative shapes showing vortex shedding, as the flow rotates and stretches around the object. Movies of the time-dependent flow can be seen on the \FlowBench~\href{https://baskargroup.bitbucket.io/#/problems}{website}.

For the simulations we performed in 2D, we ensure the fluid mesh's resolution is fine enough for the Direct Numerical Simulation (DNS) simulations. DNS is an expensive high-fidelity computational approach that solves the full Navier-Stokes equations, resolving all scales of motion in a fluid without additional turbulence modeling. The Kolmogorov scale is crucial in turbulence theory as it represents the smallest length scale that must be resolved in DNS to capture all aspects of turbulence accurately. This scale can be expressed in terms of the Reynolds number as \(\eta \sim \frac{L}{Re^{3/4}}\), where \(L\) is the characteristic large-eddy length scale. In our 2D simulation, we ensure that the mesh size is smaller than the Kolmogorov scale, calculated based on the Reynolds number, to capture the full range of turbulent motions.

The 3D LDC cases are similarly defined and simulated. Due to the computational expense of performing DNS on 3D geometries, we report 500 simulations of complex objects. Here, we focused on parametric shapes, specifically ellipsoids and tori, exhibiting various aspect ratios and orientations.

\subsection{Simulation Framework and Compute effort}\label{subsec:ComputeEfforts}
Our simulation framework is a highly parallel octree-based CFD and multiphysics code. We account for complex geometries using a robust variant of the immersed boundary method, called the Shifted Boundary Method (SBM)~\citep{main2018shifted,Main2018TheSB}. For readers interested in our framework's SBM formulation and implementation, please refer to~\citep{yang2024optimal}. Our octree-based framework has been extensively validated and used in various applications like industrial-scale CFD simulations~\citep{saurabh2021industrial}, two-phase flow~\citep{khanwale2023projection},  buoyancy-driven flows~\citep{xu2019residual}, and coupled multiphysics application~\citep{kim2024direct}. For additional validation of the incompressible flow and thermal incompressible flow solving using SBM, please refer to \secref{subsec:validations} in the appendix. 
All the 2D LDC cases are simulated using a uniform mesh of $512 \times 512$. All the 2D FPO cases are simulated using an adaptive mesh that resolved to the Kolmogorov length scale close to the object. All the 3D LDC cases are simulated using an adaptive mesh that resolved to $2\times$ the Kolmogorov scale. Illustrations of the mesh are provided in \secref{app:mesh}. We deployed this framework on one of the largest academic supercomputing clusters in the US, called TACC \Frontera~\citep{stanzione2020frontera}. These simulations required about 65K nodehours of runtime.



\subsection{MetaData}\label{subsec:meta}
\textbf{Input Fields}: We have provisioned the following input fields:
\begin{enumerate}[left=0pt,topsep=0pt,itemsep=0pt]
\item \textbf{Reynolds Number}, \textbf{Grashof Number}: We feed the Reynolds and Grashof numbers as a concordant (matching the dimensions of the flow domain e.g. for 2D lid driven cavity problem, $512\times512$) array comprising of a single integer value everwhere. 
\item \textbf{Geometry Mask}: A geometry mask ($g$) is a binary representation of a shape or object within a spatial domain. Each element in the mask can have one of two values, typically 0 or 1, where: 0 indicates that the point is outside the object and 1 indicates that the point is inside the object. The geometry mask helps in identifying and isolating the region of interest (the object) from the background or surrounding space. This information is again packaged as a concordant array with each entry marked as 0 or 1.
\item \textbf{Signed Distance Field (SDF)}: A Signed Distance Field ($s$) is a scalar field that represents the shortest distance from any point in space to the surface of a given shape. The "signed" aspect of the signed distance field indicates whether the point is inside or outside the shape. A negative value indicates the point is inside the shape, while a positive value indicates the point is outside the shape. A value of zero indicates the point is exactly on the surface of the shape. The SDF provides additional information through the distance values, which is important for understanding the spatial relationship between locations and the geometric boundary. This method offers richer geometric information compared to the binary geometry mask, as it includes both positional and distance information. Yet again, this information is packaged as a concordant array with each entry representing the nearest distance to the geometrical shape.

\end{enumerate}

\textbf{Output Fields}: We are interested in obtaining field solutions (i.e., solutions at  every point in the interior of the domain) for certain cardinal fields. For a 2D solution domain, these are: $u$ - velocity in $x$ direction, $v$ - velocity in $y$ direction, $p$ - pressure and $\theta$ - temperature. Additionally, depending on whether we are solving a steady-state or a time-dependent problem, we would have either have one final snapshot of these cardinal fields or a sequence of these fields distributed uniformly over time.

\textbf{Resolution}: We perform simulations at DNS resolution. \FlowBench{}  contains postprocessed simulation results at three different resolutions. This serves multiple purposes: First, it allows the community to systematically explore tradeoffs in the amount of data vs resolution vs accuracy. Second, since we perform non-interpolatory sub-sampling, this data allows the community to systematically test PDE super-resolution approaches. Finally, the lower-resolution datasets offer easier opportunities to train the data. For instance, at the time of submission, we could not train the 3D SciML models at the highest resolution available. For the steady state cases, we provide data at resolutions $512\times512$, $256\times256$, $128\times128$ ($\times 128$ - for 3D LDC NS). For the time dependant cases, we provide raw data at resolutions, $1024 \times 256$, and $512 \times 128$. We provide 240 snapshots for each time-dependant case. With only a light amount of postprocessing, the end user can choose from these 240 timesteps to create more managable datasets for machine learning, e.g., taking snapshots every sixth time step.

\textbf{Dataset Format}: All of our datasets are provided as numpy compressed (\texttt{.npz}) files. For each steady-state problem, we provide two \texttt{.npz} files - one for input and one for output. The input files are suffixed with the marker \texttt{"\_X.npz"}, and similarly, the corresponding output files are suffixed with the marker \texttt{"\_Y.npz"}. In 2D, each of these \texttt{.npz} files contains a 4D \texttt{numpy} tensor of the following form:
{\small
\[
    [\mathbf{samples}][\mathbf{number\_of\_channels}][\mathbf{resolution\_x}][\mathbf{resolution\_y}]
\]
}

Similarly, in 3D, each of these \texttt{.npz} files contains a 5D \texttt{numpy} tensor of the following form:
{\small
\[
    [\mathbf{samples}][\mathbf{number\_of\_channels}][\mathbf{resolution\_x}][\mathbf{resolution\_y}][\mathbf{resolution\_z}]
\]
}

For time-dependent problems, we provide a single file for each problem. This decision was taken to allow maximum flexibility to the end user in deciding what and how many time steps they want to use to train their models, as these time-dependent problems often take the shape of sequence-to-sequence formulations. In 2D, the resulting \texttt{.npz} files take the following form:
{\small
\[
    [\mathbf{samples}][\mathbf{number\_of\_time\_steps}][\mathbf{number\_of\_channels}][\mathbf{resolution\_x}][\mathbf{resolution\_y}]
\]}
We provide a total of over 10K samples spread across four families of datasets. \tabref{tab:InputOutputTensors} in the Appendix provides a detailed formulaic description of the packaging of the input and output \texttt{numpy} tensors for each of these four families.

\subsection{Evaluation Metrics and Test Datasets}\label{subsec:metrics}

\textbf{Errors}: We compute the following errors to evaluate the accuracy of all the models trained on \FlowBench{}:
\begin{itemize}[topsep=0pt, itemsep=0pt, left=0pt]
    \item \textit{Mean Squared Error}:
    \[
        \sum_{i = 1}^{n}(y_i - y_i^{'})^{2}
    \]
    \item $L_2$ \textit{Relative Error}:
    \[
        L_2 = \sqrt{\frac{\sum_{i = 1}^{n}(y_i - y_i^{'})^{2}}{\sum_{i = 1}^{n} y_i^{'2}}}
    \]
    \item $L_{\infty}$ \textit{Relative Error}:
    \[
        L_\infty  = \frac{\max \left| y_i - y_i^{'} \right|}{\max \left| y_i^{'} \right|}
    \]
\end{itemize}

\textit{Note:} In all the above definitions, $y, y^{'}$ and $n$ indicate the predicted value from a model, the ground truth and the test dataset size respectively.

\textbf{Evaluation metrics}: We recommend a hierarchy of metrics ($M1, M2, M3, M4$) to comprehensively assess the performance of trained models using \FlowBench{}: 
\begin{itemize}[topsep=0pt, itemsep=0pt, left=0pt] 
    \item \textit{$M1$: Global metrics}: Computing all three errors over the entire domain is a good primary metric to measure the accuracy of predicted velocity, pressure, and temperature fields, reflecting how closely the model's predictions match the true values.
    \item \textit{$M2$: Boundary layer metrics}: The errors in velocity, pressure, and temperature fields over a narrow region around the object. We define the boundary layer by considering the solution conditioned on the Signed Distance Field ($0 \leq SDF \leq 0.2$). This tougher metric provides insight into the model's accuracy in predicting near-surface phenomena, which are crucial for applications including flow diagnostics, shape design, and dynamic control.
    \item \textit{$M3$: Property metrics}: An application-driven metric is to evaluate the accuracy of the summary statistics -- $C_D, C_L, Nu$. The coefficients of lift and drag are critical for assessing the forces acting on the object, while the average Nusselt number predicts heat transfer rates, all of which are key for flow-thermal management and engineering applications. This metric represents spatially averaged properties. 
    \item \textit{$M4$: Residual metrics}: One can also evaluate how well the predicted field satisfies the underlying (set of) PDE. Evaluating how well continuity ($\nabla \cdot u = 0$) is satisfied all over the domain is a measure of respecting the conservation of mass. Similarly, the global average of the PDE residual evaluates how well the model satisfies the underlying PDE in the domain.
\end{itemize}

\textbf{Test Datasets}: We recommend evaluating trained models on two test scenarios
\begin{itemize}[topsep=0pt, itemsep=0pt, left=0pt] 
    \item \textit{\textbf{Standard}}: This is the standard 80-20 random split of the \FlowBench{}  data.
    \item \textit{\textbf{Hard}}: Here, we recommend splitting the data based on their operating characteristics, i.e. Reynolds number and Grashof (or Richardson) number. Samples with $(Re, Gr)$ in the intermediate range (say, the middle 70\%) should be used for training and tested against the out-of-distribution $(Re, Gr)$ samples. While the specifics of the flow patterns in this test dataset may differ from the training set, the global features remain the same, thus providing a good test of generalizability. 
\end{itemize}

Together, these metrics and datasets provide a comprehensive evaluation framework, allowing practitioners to evaluate model accuracy, physical consistency, and practical reliability.


\section{Experiments} \label{sec:experiments}

We report baseline results for training a suite of the most common neural PDE solvers. We studied the following Neural Operators and Foundation Models, reporting results on the 2D LDC-NS: (a) Fourier Neural Operator (FNO) \cite{li2021}, (b) Convolutional Neural Operators (CNO) \cite{raonic2023}, (c) DeepONet \cite{lu2020}, (d) Geometric DeepONet \cite{he2024}, (e) Wavelet Neural Operator \cite{TRIPURA2023115783}, (f) scOT \cite{herde2024poseidon}, (g) Poseidon \cite{herde2024poseidon}. Plots of field solutions for $u$, $v$, and $p$, as well as training and validation losses, are available on our \href{https://baskargroup.bitbucket.io/#/experimentalresults}{website}. For training, we adhered closely to the published code examples. All the aforementioned models were trained on a single A100 80GB GPU using the Adam optimizer with a learning rate of $10^{-3}$ and were run for 400 epochs. The validation loss for all models converged and stabilized by 400 epochs. 

\tabref{tab:mse-errors-3} through \tabref{tab:linf-errors} present a comprehensive evaluation of the performance of all the trained models. The models were assessed using the four metrics outlined in \secref{subsec:metrics}, across each geometry class and at the highest resolution ($512 \times 512$) for all seven models. Our findings indicate that in terms of both \textit{mean squared error} and $L_{\infty}$ error, the foundation models consistently outperformed the neural operators across all geometries and metrics. For the $L_{2}$ error, the foundation models showed better accuracy in all \textbf{\textit{Hard}} cases, whereas the neural operators performed comparably to the foundation models in the \textbf{\textit{Easy}} cases. The higher accuracy, especially for the \textbf{\textit{Hard}} cases, of the foundation models may be attributed to their extensive pretraining on a diverse set of PDEs conducted \textit{a priori}. We encourage further exploration and validation of these findings by the research community.

\begin{table}
\centering
\small
\setlength\extrarowheight{2pt}
\caption{The mean squared errors of various neural operators trained on the 2D LDC dataset at two different difficulty levels (Easy and Hard). All errors are reported on the testing dataset.}
\label{tab:mse-errors-3}
\begin{tabular}{|c|c|c|c|c|c|}
\hline
 & \textbf{Geometry} & \multicolumn{2}{c|}{\textbf{Easy}} & \multicolumn{2}{c|}{\textbf{Hard}} \\ \hline
\textbf{Model} &  & M1 & M2 & M1 & M2 \\ \hline

\multirow{3}{*}{\textbf{FNO}} 
& G1 & $5.8 \times 10^{-3}$ & $7.1 \times 10^{-4}$ & $9.0 \times 10^{-2}$ & $1.2 \times 10^{-2}$ \\ \cline{2-6}
& G2 & $6.8 \times 10^{-3}$ & $4.9 \times 10^{-4}$ & $1.1 \times 10^{-1}$ & $1.3 \times 10^{-2}$ \\ \cline{2-6}
& G3 & $9.7 \times 10^{-3}$ & $1.2 \times 10^{-3}$ & $1.1 \times 10^{-1}$ & $1.9 \times 10^{-2}$ \\ \hline

\multirow{3}{*}{\textbf{CNO}} 
& G1 & $3.1 \times 10^{-3}$ & $7.4 \times 10^{-4}$ & $2.9 \times 10^{-2}$ & $3.8 \times 10^{-3}$ \\ \cline{2-6}
& G2 & $2.4 \times 10^{-3}$ & $7.8 \times 10^{-4}$ & $3.3 \times 10^{-2}$ & $3.9 \times 10^{-3}$ \\ \cline{2-6}
& G3 & $3.1 \times 10^{-3}$ & $7.3 \times 10^{-4}$ & $3.8 \times 10^{-2}$ & $6.1 \times 10^{-3}$ \\ \hline

\multirow{3}{*}{\textbf{DeepONet}} 
& G1 & $4.9 \times 10^{-3}$ & $1.8 \times 10^{-3}$ & $7.8 \times 10^{-2}$ & $1.9 \times 10^{-2}$ \\ \cline{2-6}
& G2 & $2.3 \times 10^{-3}$ & $7.0 \times 10^{-4}$ & $8.8 \times 10^{-2}$ & $1.5 \times 10^{-2}$ \\ \cline{2-6}
& G3 & $6.0 \times 10^{-3}$ & $1.4 \times 10^{-3}$ & $8.6 \times 10^{-2}$ & $2.2 \times 10^{-2}$ \\ \hline

\multirow{3}{*}{\textbf{Geom. DeepONet}} 
& G1 & $1.1 \times 10^{-3}$ & $2.4 \times 10^{-4}$ & $5.5 \times 10^{-2}$ & $7.8 \times 10^{-3}$ \\ \cline{2-6}
& G2 & $9.3 \times 10^{-4}$ & $1.2 \times 10^{-4}$ & $4.2 \times 10^{-2}$ & $4.6 \times 10^{-3}$ \\ \cline{2-6}
& G3 & $2.6 \times 10^{-3}$ & $6.7 \times 10^{-4}$ & $6.4 \times 10^{-2}$ & $1.1 \times 10^{-2}$ \\ \hline

\multirow{3}{*}{\textbf{WNO}} 
& G1 & $6.2 \times 10^{-3}$ & $1.9 \times 10^{-3}$ & $9.8 \times 10^{-2}$ & $1.8 \times 10^{-2}$ \\ \cline{2-6}
& G2 & $2.9 \times 10^{-3}$ & $6.5 \times 10^{-4}$ & $8.2 \times 10^{-2}$ & $1.3 \times 10^{-2}$ \\ \cline{2-6}
& G3 & $9.8 \times 10^{-2}$ & $1.8 \times 10^{-2}$ & $9.2 \times 10^{-2}$ & $2.0 \times 10^{-2}$ \\ \hline

\multirow{3}{*}{\textbf{scot}} 
& G1 & $5.5 \times 10^{-4}$ & $1.7 \times 10^{-4}$ & $3.6 \times 10^{-2}$ & $3.1 \times 10^{-3}$ \\ \cline{2-6}
& G2 & $4.1 \times 10^{-4}$ & $1.3 \times 10^{-4}$ & $4.1 \times 10^{-2}$ & $5.1 \times 10^{-3}$ \\ \cline{2-6}
& G3 & $1.3 \times 10^{-3}$ & $5.2 \times 10^{-4}$ & $1.7 \times 10^{-2}$ & $1.0 \times 10^{-2}$ \\ \hline

\multirow{3}{*}{\textbf{poseidon}} 
& G1 & $2.2 \times 10^{-4}$ & $7.7 \times 10^{-5}$ & $1.9 \times 10^{-2}$ & $\mathbf{1.6 \times 10^{-3}}$ \\ \cline{2-6}
& G2 & $\mathbf{1.4 \times 10^{-4}}$ & $\mathbf{4.9 \times 10^{-5}}$ & $1.8 \times 10^{-2}$ & $2.2 \times 10^{-3}$ \\ \cline{2-6}
& G3 & $4.0 \times 10^{-4}$ & $1.4 \times 10^{-4}$ & $\mathbf{1.7 \times 10^{-2}}$ & $3.3 \times 10^{-3}$ \\ \hline

\end{tabular}
\end{table}

\small{

\begin{table}
\centering
\small
\setlength\extrarowheight{2pt}
\caption{The relative \(L_2\) error of various neural operators trained on the 2D LDC dataset at two different difficulty levels (Easy and Hard). All metrics are reported on the testing dataset.}
\label{tab:metrics-errors}
\begin{tabular}{|c|c|c|c|c|c|}
\hline
 & \textbf{Geometry} & \multicolumn{2}{c|}{\textbf{Easy}} & \multicolumn{2}{c|}{\textbf{Hard}} \\ \hline
\textbf{Model} &  & M1 & M2 & M1 & M2 \\ \hline

\multirow{3}{*}{\textbf{FNO}} 
& G1 & $3.3 \times 10^{-1}$ & $3.6 \times 10^{-1}$ & $4.9 \times 10^{-1}$ & $5.9 \times 10^{-1}$ \\ \cline{2-6}
& G2 & $3.8 \times 10^{-1}$ & $4.3 \times 10^{-1}$ & $5.3 \times 10^{-1}$ & $5.4 \times 10^{-1}$ \\ \cline{2-6}
& G3 & $4.9 \times 10^{-1}$ & $6.0 \times 10^{-1}$ & $7.4 \times 10^{-1}$ & $9.3 \times 10^{-1}$ \\ \hline

\multirow{3}{*}{\textbf{CNO}} 
& G1 & $2.7 \times 10^{-1}$ & $4.3 \times 10^{-1}$ & $3.4 \times 10^{-1}$ & $5.5 \times 10^{-1}$ \\ \cline{2-6}
& G2 & $2.4 \times 10^{-1}$ & $3.9 \times 10^{-1}$ & $4.3 \times 10^{-1}$ & $7.3 \times 10^{-1}$ \\ \cline{2-6}
& G3 & $4.9 \times 10^{-1}$ & $5.4 \times 10^{-1}$ & $6.3 \times 10^{-1}$ & $7.9 \times 10^{-1}$ \\ \hline

\multirow{3}{*}{\textbf{DeepONet}} 
& G1 & $3.9 \times 10^{-1}$ & $5.1 \times 10^{-1}$ & $5.9 \times 10^{-1}$ & $7.2 \times 10^{-1}$ \\ \cline{2-6}
& G2 & $4.9 \times 10^{-1}$ & $5.7 \times 10^{-1}$ & $6.4 \times 10^{-1}$ & $8.9 \times 10^{-1}$ \\ \cline{2-6}
& G3 & $6.2 \times 10^{-1}$ & $7.1 \times 10^{-1}$ & $8.1 \times 10^{-1}$ & $1.0 \times 10^0$ \\ \hline

\multirow{3}{*}{\textbf{Geom. DeepONet}} 
& G1 & $\mathbf{1.3 \times 10^{-1}}$ & $2.5 \times 10^{-1}$ & $3.8 \times 10^{-1}$ & $4.4 \times 10^{-1}$ \\ \cline{2-6}
& G2 & $1.4 \times 10^{-1}$ & $\mathbf{2.3 \times 10^{-1}}$ & $4.4 \times 10^{-1}$ & $4.9 \times 10^{-1}$ \\ \cline{2-6}
& G3 & $3.9 \times 10^{-1}$ & $4.7 \times 10^{-1}$ & $6.1 \times 10^{-1}$ & $6.8 \times 10^{-1}$ \\ \hline

\multirow{3}{*}{\textbf{WNO}} 
& G1 & $2.7 \times 10^{-1}$ & $4.1 \times 10^{-1}$ & $4.6 \times 10^{-1}$ & $7.2 \times 10^{-1}$ \\ \cline{2-6}
& G2 & $3.4 \times 10^{-1}$ & $3.8 \times 10^{-1}$ & $4.3 \times 10^{-1}$ & $6.7 \times 10^{-1}$ \\ \cline{2-6}
& G3 & $6.8 \times 10^{-1}$ & $8.2 \times 10^{-1}$ & $9.2 \times 10^{-1}$ & $1.1 \times 10^{0}$ \\ \hline

\multirow{3}{*}{\textbf{scot}} 
& G1 & $2.4 \times 10^{-1}$ & $3.1 \times 10^{-1}$ & $3.6 \times 10^{-1}$ & $4.4 \times 10^{-1}$ \\ \cline{2-6}
& G2 & $1.8 \times 10^{-1}$ & $2.9 \times 10^{-1}$ & $4.4 \times 10^{-1}$ & $5.7 \times 10^{-1}$ \\ \cline{2-6}
& G3 & $2.1 \times 10^{-1}$ & $2.8 \times 10^{-1}$ & $5.4 \times 10^{-1}$ & $7.4 \times 10^{-1}$ \\ \hline

\multirow{3}{*}{\textbf{poseidon}} 
& G1 & $1.9 \times 10^{-1}$ & $2.7 \times 10^{-1}$ & $\mathbf{3.2 \times 10^{-1}}$ & $\mathbf{3.6 \times 10^{-1}}$ \\ \cline{2-6}
& G2 & $2.4 \times 10^{-1}$ & $2.7 \times 10^{-1}$ & $\mathbf{3.2 \times 10^{-1}}$ & $4.2 \times 10^{-1}$ \\ \cline{2-6}
& G3 & $3.9 \times 10^{-1}$ & $4.9 \times 10^{-1}$ & $5.2 \times 10^{-1}$ & $6.4 \times 10^{-1}$ \\ \hline

\end{tabular}
\end{table}

\small{

\begin{table}
\centering
\small
\setlength\extrarowheight{2pt}
\caption{The relative \(L_\infty\) error of various neural operators trained on the 2D LDC dataset at two different difficulty levels (Easy and Hard). All metrics are reported on the testing dataset.}
\label{tab:linf-errors}
\begin{tabular}{|c|c|c|c|c|c|}
\hline
 & \textbf{Geometry} & \multicolumn{2}{c|}{\textbf{Easy}} & \multicolumn{2}{c|}{\textbf{Hard}} \\ \hline
\textbf{Model} &  & M1 & M2 & M1 & M2 \\ \hline

\multirow{3}{*}{\textbf{FNO}} 
& G1 & $6.1 \times 10^{-1}$ & $7.2 \times 10^{-1}$ & $8.4 \times 10^{-1}$ & $9.2 \times 10^{-1}$ \\ \cline{2-6}
& G2 & $7.4 \times 10^{-1}$ & $8.1 \times 10^{-1}$ & $9.1 \times 10^{-1}$ & $9.7 \times 10^{-1}$ \\ \cline{2-6}
& G3 & $9.8 \times 10^{-1}$ & $1.1 \times 10^0$ & $1.2 \times 10^0$ & $1.3 \times 10^0$ \\ \hline

\multirow{3}{*}{\textbf{CNO}} 
& G1 & $4.9 \times 10^{-1}$ & $7.3 \times 10^{-1}$ & $\mathbf{5.7 \times 10^{-1}}$ & $8.3 \times 10^{-1}$ \\ \cline{2-6}
& G2 & $5.4 \times 10^{-1}$ & $8.0 \times 10^{-1}$ & $7.2 \times 10^{-1}$ & $9.6 \times 10^{-1}$ \\ \cline{2-6}
& G3 & $9.3 \times 10^{-1}$ & $1.1 \times 10^0$ & $1.1 \times 10^0$ & $1.3 \times 10^0$ \\ \hline

\multirow{3}{*}{\textbf{DeepONet}} 
& G1 & $8.3 \times 10^{-1}$ & $9.1 \times 10^{-1}$ & $9.9 \times 10^{-1}$ & $1.1 \times 10^0$ \\ \cline{2-6}
& G2 & $9.0 \times 10^{-1}$ & $1.0 \times 10^0$ & $1.1 \times 10^0$ & $1.2 \times 10^0$ \\ \cline{2-6}
& G3 & $1.1 \times 10^0$ & $1.3 \times 10^0$ & $1.3 \times 10^0$ & $1.4 \times 10^0$ \\ \hline

\multirow{3}{*}{\textbf{Geom. DeepONet}} 
& G1 & $\mathbf{3.2 \times 10^{-1}}$ & $5.3 \times 10^{-1}$ & $6.8 \times 10^{-1}$ & $8.4 \times 10^{-1}$ \\ \cline{2-6}
& G2 & $3.3 \times 10^{-1}$ & $5.1 \times 10^{-1}$ & $8.4 \times 10^{-1}$ & $9.2 \times 10^{-1}$ \\ \cline{2-6}
& G3 & $9.4 \times 10^{-1}$ & $1.1 \times 10^0$ & $1.1 \times 10^0$ & $1.2 \times 10^0$ \\ \hline

\multirow{3}{*}{\textbf{WNO}} 
& G1 & $6.2 \times 10^{-1}$ & $9.1 \times 10^{-1}$ & $8.7 \times 10^{-1}$ & $1.1 \times 10^0$ \\ \cline{2-6}
& G2 & $7.1 \times 10^{-1}$ & $9.7 \times 10^{-1}$ & $1.1 \times 10^0$ & $1.2 \times 10^0$ \\ \cline{2-6}
& G3 & $1.3 \times 10^0$ & $1.5 \times 10^0$ & $1.4 \times 10^0$ & $1.6 \times 10^0$ \\ \hline

\multirow{3}{*}{\textbf{scot}} 
& G1 & $2.7 \times 10^{-1}$ & $5.3 \times 10^{-1}$ & $6.2 \times 10^{-1}$ & $7.4 \times 10^{-1}$ \\ \cline{2-6}
& G2 & $3.4 \times 10^{-1}$ & $5.7 \times 10^{-1}$ & $6.9 \times 10^{-1}$ & $7.9 \times 10^{-1}$ \\ \cline{2-6}
& G3 & $7.1 \times 10^{-1}$ & $8.2 \times 10^{-1}$ & $9.1 \times 10^{-1}$ & $1.1 \times 10^0$ \\ \hline

\multirow{3}{*}{\textbf{poseidon}} 
& G1 & $3.3 \times 10^{-1}$ & $\mathbf{4.3 \times 10^{-1}}$ & $\mathbf{5.7 \times 10^{-1}}$ & $\mathbf{6.2 \times 10^{-1}}$ \\ \cline{2-6}
& G2 & $4.1 \times 10^{-1}$ & $5.7 \times 10^{-1}$ & $7.1 \times 10^{-1}$ & $7.8 \times 10^{-1}$ \\ \cline{2-6}
& G3 & $6.4 \times 10^{-1}$ & $7.7 \times 10^{-1}$ & $9.3 \times 10^{-1}$ & $1.1 \times 10^0$ \\ \hline

\end{tabular}
\end{table}

\begin{table}
\centering
\small
\setlength\extrarowheight{2pt}
\caption{The mean squared errors of various neural operators trained on the 2D LDC dataset at two different difficulty levels (Easy and Hard). All metrics are reported on the testing dataset.}
\label{tab:mse-errors-3-m3}
\begin{tabular}{|c|c|c|c|}
\hline
 & \textbf{Geometry} & \multicolumn{1}{c|}{\textbf{Easy}} & \multicolumn{1}{c|}{\textbf{Hard}} \\ \hline
\textbf{Model} &  & M3 & M3 \\ \hline

\multirow{3}{*}{\textbf{FNO}} 
& G1 & $2.7 \times 10^{-3}$ & $4.6 \times 10^{-1}$ \\ \cline{2-4}
& G2 & $2.9 \times 10^{-3}$ & $4.2 \times 10^{-1}$ \\ \cline{2-4}
& G3 & $1.3 \times 10^{-2}$ & $3.0 \times 10^{-1}$ \\ \hline

\multirow{3}{*}{\textbf{CNO}} 
& G1 & $2.1 \times 10^{-2}$ & $1.2 \times 10^{-1}$ \\ \cline{2-4}
& G2 & $8.4 \times 10^{-3}$ & $8.4 \times 10^{-2}$ \\ \cline{2-4}
& G3 & $3.1 \times 10^{-2}$ & $1.6 \times 10^{-1}$ \\ \hline

\multirow{3}{*}{\textbf{DeepONet}} 
& G1 & $1.8 \times 10^{-2}$ & $6.5 \times 10^{-1}$ \\ \cline{2-4}
& G2 & $7.0 \times 10^{-3}$ & $4.8 \times 10^{-1}$ \\ \cline{2-4}
& G3 & $3.7 \times 10^{-2}$ & $4.6 \times 10^{-1}$ \\ \hline

\multirow{3}{*}{\textbf{Geom. DeepONet}} 
& G1 & $2.2 \times 10^{-3}$ & $4.4 \times 10^{-1}$ \\ \cline{2-4}
& G2 & $2.6 \times 10^{-3}$ & $2.6 \times 10^{-1}$ \\ \cline{2-4}
& G3 & $2.0 \times 10^{-2}$ & $2.8 \times 10^{-1}$ \\ \hline

\multirow{3}{*}{\textbf{WNO}} 
& G1 & $2.1 \times 10^{-2}$ & $7.1 \times 10^{-1}$ \\ \cline{2-4}
& G2 & $8.7 \times 10^{-3}$ & $4.5 \times 10^{-1}$ \\ \cline{2-4}
& G3 & $7.1 \times 10^{-1}$ & $5.5 \times 10^{-1}$ \\ \hline

\multirow{3}{*}{\textbf{scot}} 
& G1 & $1.9 \times 10^{-3}$ & $3.1 \times 10^{-1}$ \\ \cline{2-4}
& G2 & $2.5 \times 10^{-3}$ & $1.3 \times 10^{-1}$ \\ \cline{2-4}
& G3 & $1.7 \times 10^{-2}$ & $1.9 \times 10^{-1}$ \\ \hline

\multirow{3}{*}{\textbf{poseidon}} 
& G1 & $5.2 \times 10^{-4}$ & $1.4 \times 10^{-1}$ \\ \cline{2-4}
& G2 & $\mathbf{4.1 \times 10^{-4}}$ & $\mathbf{6.2 \times 10^{-2}}$ \\ \cline{2-4}
& G3 & $5.4 \times 10^{-3}$ & $7.0 \times 10^{-2}$ \\ \hline

\end{tabular}
\end{table}

\begin{table}
\centering
\small
\setlength\extrarowheight{2pt}
\caption{The relative \(L_2\) error of various neural operators trained on the 2D LDC dataset at two different difficulty levels (Easy and Hard). All metrics are reported on the testing dataset.}
\label{tab:metrics-errors-l2-m3}
\begin{tabular}{|c|c|c|c|}
\hline
 & \textbf{Geometry} & \multicolumn{1}{c|}{\textbf{Easy}} & \multicolumn{1}{c|}{\textbf{Hard}} \\ \hline
\textbf{Model} &  & M3 & M3 \\ \hline

\multirow{3}{*}{\textbf{FNO}} 
& G1 & $7.8 \times 10^{-1}$ & $8.2 \times 10^{-1}$ \\ \cline{2-4}
& G2 & $6.8 \times 10^{-1}$ & $3.9 \times 10^{0}$ \\ \cline{2-4}
& G3 & $8.7 \times 10^{-1}$ & $2.2 \times 10^{0}$ \\ \hline

\multirow{3}{*}{\textbf{CNO}} 
& G1 & $1.4 \times 10^{0}$ & $1.2 \times 10^{0}$ \\ \cline{2-4}
& G2 & $1.0 \times 10^{0}$ & $3.6 \times 10^{0}$ \\ \cline{2-4}
& G3 & $1.6 \times 10^{0}$ & $5.3 \times 10^{0}$ \\ \hline

\multirow{3}{*}{\textbf{DeepONet}} 
& G1 & $1.4 \times 10^{0}$ & $1.3 \times 10^{0}$ \\ \cline{2-4}
& G2 & $8.4 \times 10^{-1}$ & $2.0 \times 10^{0}$ \\ \cline{2-4}
& G3 & $1.4 \times 10^{0}$ & $2.6 \times 10^{0}$ \\ \hline

\multirow{3}{*}{\textbf{Geom. DeepONet}} 
& G1 & $6.1 \times 10^{-1}$ & $1.2 \times 10^{0}$ \\ \cline{2-4}
& G2 & $6.0 \times 10^{-1}$ & $1.8 \times 10^{0}$ \\ \cline{2-4}
& G3 & $8.4 \times 10^{-1}$ & $3.2 \times 10^{0}$ \\ \hline

\multirow{3}{*}{\textbf{WNO}} 
& G1 & $1.9 \times 10^{0}$ & $2.5 \times 10^{0}$ \\ \cline{2-4}
& G2 & $1.5 \times 10^{0}$ & $4.8 \times 10^{0}$ \\ \cline{2-4}
& G3 & $2.5 \times 10^{0}$ & $7.8 \times 10^{0}$ \\ \hline

\multirow{3}{*}{\textbf{scot}} 
& G1 & $4.8 \times 10^{-1}$ & $1.0 \times 10^{0}$ \\ \cline{2-4}
& G2 & $7.6 \times 10^{-1}$ & $2.1 \times 10^{0}$ \\ \cline{2-4}
& G3 & $1.1 \times 10^{0}$ & $1.2 \times 10^{0}$ \\ \hline

\multirow{3}{*}{\textbf{poseidon}} 
& G1 & $\mathbf{2.5 \times 10^{-1}}$ & $\mathbf{5.3 \times 10^{-1}}$ \\ \cline{2-4}
& G2 & $3.7 \times 10^{-1}$ & $9.7 \times 10^{-1}$ \\ \cline{2-4}
& G3 & $3.7 \times 10^{-1}$ & $8.3 \times 10^{-1}$ \\ \hline

\end{tabular}
\end{table}

\begin{table}
\centering
\small
\setlength\extrarowheight{2pt}
\caption{The relative \(L_{\infty}\) error of various neural operators trained on the 2D LDC dataset at two different difficulty levels (Easy and Hard). All metrics are reported on the testing dataset.}
\label{tab:metrics-errors-linf-m3}
\begin{tabular}{|c|c|c|c|}
\hline
 & \textbf{Geometry} & \multicolumn{1}{c|}{\textbf{Easy}} & \multicolumn{1}{c|}{\textbf{Hard}} \\ \hline
\textbf{Model} &  & M3 & M3 \\ \hline

\multirow{3}{*}{\textbf{FNO}} 
& G1 & $9.8 \times 10^{-1}$ & $6.1 \times 10^{0}$ \\ \cline{2-4}
& G2 & $8.5 \times 10^{-1}$ & $5.2 \times 10^{0}$ \\ \cline{2-4}
& G3 & $1.2 \times 10^{0}$ & $4.8 \times 10^{0}$ \\ \hline

\multirow{3}{*}{\textbf{CNO}} 
& G1 & $3.5 \times 10^{0}$ & $4.0 \times 10^{0}$ \\ \cline{2-4}
& G2 & $2.1 \times 10^{0}$ & $7.1 \times 10^{0}$ \\ \cline{2-4}
& G3 & $4.7 \times 10^{0}$ & $8.3 \times 10^{0}$ \\ \hline

\multirow{3}{*}{\textbf{DeepONet}} 
& G1 & $3.3 \times 10^{0}$ & $1.2 \times 10^{1}$ \\ \cline{2-4}
& G2 & $2.6 \times 10^{0}$ & $9.2 \times 10^{0}$ \\ \cline{2-4}
& G3 & $3.1 \times 10^{0}$ & $7.6 \times 10^{0}$ \\ \hline

\multirow{3}{*}{\textbf{Geom. DeepONet}} 
& G1 & $1.1 \times 10^{0}$ & $3.2 \times 10^{0}$ \\ \cline{2-4}
& G2 & $1.1 \times 10^{0}$ & $5.4 \times 10^{0}$ \\ \cline{2-4}
& G3 & $1.5 \times 10^{0}$ & $7.0 \times 10^{0}$ \\ \hline

\multirow{3}{*}{\textbf{WNO}} 
& G1 & $3.9 \times 10^{0}$ & $1.2 \times 10^{1}$ \\ \cline{2-4}
& G2 & $3.2 \times 10^{0}$ & $7.1 \times 10^{0}$ \\ \cline{2-4}
& G3 & $5.2 \times 10^{0}$ & $1.6 \times 10^{1}$ \\ \hline

\multirow{3}{*}{\textbf{scot}} 
& G1 & $7.7 \times 10^{-1}$ & $3.2 \times 10^{0}$ \\ \cline{2-4}
& G2 & $1.3 \times 10^{0}$ & $5.3 \times 10^{0}$ \\ \cline{2-4}
& G3 & $1.9 \times 10^{0}$ & $3.1 \times 10^{0}$ \\ \hline

\multirow{3}{*}{\textbf{poseidon}} 
& G1 & $\mathbf{2.9 \times 10^{-1}}$ & $\mathbf{1.6 \times 10^{0}}$ \\ \cline{2-4}
& G2 & $4.5 \times 10^{-1}$ & $3.1 \times 10^{0}$ \\ \cline{2-4}
& G3 & $5.1 \times 10^{-1}$ & $2.6 \times 10^{0}$ \\ \hline

\end{tabular}
\end{table}

\small{
\begin{table}
\centering
\small
\setlength\extrarowheight{2pt}
\caption{The residual errors (M4) of various neural operators trained on the 2D LDC dataset at two different difficulty levels (Easy and Hard). All metrics are reported on the testing dataset. Errors reported on the test dataset (per sample) normalized by the total number of mesh points.}
\label{tab:linf-errors}
\begin{tabular}{|c|c|c|c|c|c|}
\hline
 & \textbf{Geometry} & \multicolumn{2}{c|}{\textbf{Easy}} & \multicolumn{2}{c|}{\textbf{Hard}} \\ \hline
\textbf{Model} &  & \textbf{momentum} & \textbf{continuity} & \textbf{momentum} & \textbf{continuity} \\ \hline

\multirow{3}{*}{\textbf{FNO}}
& G1 & $1.03 \times 10^{-4}$ & $9.11 \times 10^{-8}$ & $8.62 \times 10^{-5}$ & $9.30 \times 10^{-8}$ \\ \cline{2-6}
& G2 & $5.80 \times 10^{-5}$ & $7.05 \times 10^{-8}$ & $7.67 \times 10^{-5}$ & $8.26 \times 10^{-8}$ \\ \cline{2-6}
& G3 & $7.62 \times 10^{-5}$ & $\mathbf{6.76 \times 10^{-8}}$ & $7.79 \times 10^{-5}$ & $\mathbf{1.55 \times 10^{-8}}$ \\ \hline

\multirow{3}{*}{\textbf{CNO}}
& G1 & $8.80 \times 10^{-5}$ & $4.36 \times 10^{-7}$ & $1.75 \times 10^{-4}$ & $1.40 \times 10^{-6}$ \\ \cline{2-6}
& G2 & $9.11 \times 10^{-5}$ & $4.01 \times 10^{-7}$ & $1.30 \times 10^{-4}$ & $6.82 \times 10^{-7}$ \\ \cline{2-6}
& G3 & $1.04 \times 10^{-4}$ & $3.46 \times 10^{-7}$ & $1.99 \times 10^{-4}$ & $1.13 \times 10^{-6}$ \\ \hline

\multirow{3}{*}{\textbf{DeepONet}}
& G1 & $5.09 \times 10^{-5}$ & $8.96 \times 10^{-8}$ & $7.02 \times 10^{-5}$ & $4.55 \times 10^{-7}$ \\ \cline{2-6}
& G2 & \textbf{$\mathbf{4.51 \times 10^{-5}}$} & $1.39 \times 10^{-7}$ & $\mathbf{5.92 \times 10^{-5}}$ & $9.35 \times 10^{-7}$ \\ \cline{2-6}
& G3 & $5.72 \times 10^{-5}$ & $5.07 \times 10^{-7}$ & $7.07 \times 10^{-5}$ & $1.92 \times 10^{-7}$ \\ \hline

\multirow{3}{*}{\textbf{Geom. DeepONet}}
& G1 & $6.32 \times 10^{-5}$ & $1.30 \times 10^{-7}$ & $8.96 \times 10^{-5}$ & $8.40 \times 10^{-7}$ \\ \cline{2-6}
& G2 & $5.22 \times 10^{-5}$ & $9.85 \times 10^{-8}$ & $1.25 \times 10^{-4}$ & $2.51 \times 10^{-7}$ \\ \cline{2-6}
& G3 & $4.55 \times 10^{-5}$ & $1.93 \times 10^{-7}$ & $9.29 \times 10^{-5}$ & $2.33 \times 10^{-7}$ \\ \hline

\multirow{3}{*}{\textbf{WNO}}
& G1 & $1.29 \times 10^{-4}$ & $9.81 \times 10^{-7}$ & $2.50 \times 10^{-4}$ & $4.00 \times 10^{-7}$ \\ \cline{2-6}
& G2 & $1.56 \times 10^{-4}$ & $2.82 \times 10^{-7}$ & $2.27 \times 10^{-4}$ & $3.73 \times 10^{-7}$ \\ \cline{2-6}
& G3 & $1.32 \times 10^{-4}$ & $9.36 \times 10^{-7}$ & $4.52 \times 10^{-4}$ & $5.52 \times 10^{-7}$ \\ \hline

\multirow{3}{*}{\textbf{scot}}
& G1 & $1.00 \times 10^{-4}$ & $1.92 \times 10^{-7}$ & $1.57 \times 10^{-4}$ & $4.39 \times 10^{-7}$ \\ \cline{2-6}
& G2 & $8.52 \times 10^{-5}$ & $6.89 \times 10^{-8}$ & $1.50 \times 10^{-4}$ & $1.96 \times 10^{-7}$ \\ \cline{2-6}
& G3 & $9.28 \times 10^{-5}$ & $2.25 \times 10^{-7}$ & $1.35 \times 10^{-4}$ & $2.54 \times 10^{-7}$ \\ \hline

\multirow{3}{*}{\textbf{poseidon}}
& G1 & $9.62 \times 10^{-5}$ & $1.35 \times 10^{-7}$ & $1.99 \times 10^{-4}$ & $1.68 \times 10^{-7}$ \\ \cline{2-6}
& G2 & $8.35 \times 10^{-5}$ & $1.39 \times 10^{-7}$ & $2.10 \times 10^{-4}$ & $5.32 \times 10^{-7}$ \\ \cline{2-6}
& G3 & $9.64 \times 10^{-5}$ & $1.46 \times 10^{-7}$ & $1.75 \times 10^{-4}$ & $2.47 \times 10^{-7}$ \\ \hline
\end{tabular}
\end{table}
}

\normalsize{
\section{Conclusions}\label{sec:limit_conclusion}
We introduce a comprehensive benchmark dataset designed for evaluating neural solvers of flow simulations over complex geometries. \FlowBench{} encompasses 2D and 3D simulations, covering many scenarios, from steady-state problems to time-dependent problems. \FlowBench{} aims to help develop scientific machine learning models by offering a complex dataset designed to challenge and benchmark their performance. Neural PDE solvers that can account for the effect of complex geometries can have a major impact on various applications ranging from bioengineering and power production to automotive and aerospace engineering defined by the interaction of complex geometrical objects with a fluid medium. 

\textbf{Limitations}: (1) Our evaluation of existing neural PDE solvers is limited to one of the four \FlowBench{} datasets and on seven neural PDE models. We encourage the community to contribute by evaluating a wider range of approaches using this dataset and the proposed metrics. Additionally, we plan to expand \FlowBench{} with more data—particularly for 3D simulations—and invite the CFD community to do the same, extending the dataset to cover higher $Re$ and $Gr$ operating conditions.
}

\section{Acknowledgements}
We gratefully acknowledge support from the NAIRR pilot program for computational access to TACC Frontera. This work is supported by the AI Research Institutes program supported by NSF and USDA-NIFA under AI Institute: for Resilient Agriculture, Award No. 2021-67021-35329. We also acknowledge partial support through NSF 2053760.

\newpage









\bibliographystyle{unsrtnat}
\bibliography{references}

\appendix
\newpage

\section{Details of the CFD simulation framework}

 Our CFD framework is a well-validated and massively scalable software suite that uses a variant of the immersed boundary method integrated with octree meshes to perform highly efficient and accurate Large-Eddy Simulations (LES) of flows around complex geometries. This framework demonstrates scalability of up to 32,000 processors, achieved through several key innovations, including (a) Rapid in-out tests: These tests quickly determine whether a point is inside or outside a given geometry, significantly speeding up the simulation process. (b) adaptive quadrature: This technique ensures accurate force evaluation by dynamically adjusting the numerical integration based on the local complexity of the geometry, and (c) Tensorized operations: These operations optimize performance by leveraging tensor algebra for computational efficiency. Additional details of the CFD framework, including implementation, are provided in~\cite {saurabh2021industrial,saurabh2023scalable}.

 Our datasets and code are licensed under CC-BY-NC-4.0. Next, we briefly detail the mathematics of the shifted boundary method. 

\subsection{Formulation of SBM for Navier Stokes}\label{sec:SBM_NS}

The Shifted Boundary Method (SBM)~\citep{main2018shifted,Main2018TheSB,atallah2020second,atallah2021shifted,yang2024optimal} is a numerical approach for solving partial differential equations (PDEs) on complex geometries without the need for body-fitted meshes. It is a robust variant of the immersed boundary methods used in CFD~\citep{peskin1972flow, zhang2004immersed,mittal2008versatile,Parvizian:07.1,kamensky2015immersogeometric,xu2021computational}. In SBM, the boundary conditions are imposed not on the actual boundary of the immersed object but on a nearby surrogate boundary. This surrogate boundary is chosen to conform to a Cartesian mesh, and the boundary conditions are corrected using Taylor expansions. This method effectively transforms the problem into one on a body-fitted domain, significantly simplifying the mesh generation process while maintaining accuracy and stability. The surrogate boundary is optimally chosen to minimize numerical errors, and the method demonstrates excellent scalability and efficiency, particularly when applied to adaptive octree meshes (see, for instance,~\citet{yang2024optimal}). This makes SBM particularly suitable for simulations involving complex geometries and multiphysics couplings. In SBM, additional terms are incorporated into the standard stabilized finite element formulations for solving the Navier-Stokes equations. These terms include a \textit{consistency} term (which appears due to integration by parts operation), an \textit{adjoint consistency} term (which is included to ensure optimal convergence rates), and the \textit{penalty} term (which ensures that as the mesh size is reduced, the boundary conditions asymptote to the true boundary conditions)  :

\begin{align}
\widetilde{B_{NS}^{VMS}} = & B_{NS}^{VMS} \underbrace{- \bigg<w_i^{c, h}, \frac{1}{Re} (\pd{u_i^{c,h}}{x_j}+\pd{u_j^{c,h}}{x_i}) \Tilde{n_j} - p^{c,h} \Tilde{n_i} \bigg>_{\tGD}}_{Consistency \; term} \nonumber\\
& \underbrace{- \bigg<\frac{1}{Re} (\pd{w_i^{c, h}}{x_j}+\pd{w_j^{c, h}}{x_i}) \Tilde{n_j} + q^{c, h} \Tilde{n_i}, u_i^{c,h} + \pd{u_i^{c,h}}{x_j}d_j \bigg>_{\tGD}}_{Adjoint \; consistency \; term}  \nonumber\\
& \underbrace{+ \frac{\beta}{h \cdot Re} \bigg<w_i^{c, h} +\pd{w_i^{c, h}}{x_j}d_j, u_i^{c,h} + \pd{u_i^{c,h}}{x_j}d_j \bigg>_{\tGD}}_{Penalty \; term},
\end{align}

and
\begin{align}
\widetilde{F_{NS}^{VMS}} = & F_{NS}^{VMS} \underbrace{- \bigg<\frac{1}{Re} (\pd{w_i^{c, h}}{x_j}+\pd{w_j^{c, h}}{x_i}) \Tilde{n_j} + q^{c, h} \Tilde{n_i}, g_i \bigg>_{\tGD}}_{Adjoint \; Consistency \; Term} \nonumber\\
&\underbrace{+ \frac{\beta}{h \cdot Re}\bigg<w_i^{c, h} +\pd{w_i^{c, h}}{x_j}d_j, g_i \bigg>_{\tGD}}_{Penalty \; Term},
\end{align}

where $g_i$ is the boundary condition we want to apply, $\Tilde{n_i}$ is the unite outward-pointing normal, $h$ is the element size, $\beta$ is the penalty parameter for the Navier-Stokes equation, $B_{NS}^{VMS}$ is the bilinear weak form for Navier-Stokes without SBM, and $F_{NS}^{VMS}$ is the linear weak for Navier-Stokes form without SBM. The formulation for Navier-Stokes with SBM can be expressed as:
\begin{align}
 \widetilde{B_{NS}^{VMS}} - \widetilde{F_{NS}^{VMS}} = 0.
\end{align}

\subsection{Formulation of SBM for Heat Transfer}\label{app:SBM_HT}

Similar to the Navier-Stokes equations, we use SBM to apply Dirichlet boundary conditions in the energy equation. Essentially, we use SBM to set the temperature to a desired value ($T_D$) on the geometries. The formulation for energy equation with SBM is:
\begin{align}
 \widetilde{B_{HT}^{VMS}} - \widetilde{F_{HT}^{VMS}} = 0,
\end{align}
with
\begin{align}
\widetilde{B_{HT}^{VMS}} = & B_{HT}^{VMS}  \underbrace{-  \frac{1}{Pe} \bigg<l^{c,h}, \pd{T^{c,h}}{x_j} \tilde{n_j} \bigg>_{\tGD}}_{\mathrm{Consistency \; term}} 
- \underbrace{ \frac{1}{Pe} \bigg< \pd{l^{c,h}}{x_j} \tilde{n_j}, T^{c,h} + \pd{T^{c,h}}{x_j}d_j \bigg>_{\tGD}}_{\mathrm{Adjoint \; consistency \; term}} \nonumber\\
& \underbrace{+  \frac{\alpha}{h \cdot Pe} \bigg<l^{c,h} +\pd{l^{c,h}}{x_j}d_j, T^{c,h} + \pd{T^{c,h}}{x_j}d_j \bigg>_{\tGD}}_{\mathrm{Penalty \; term}},
\end{align}

and
\begin{align}
\widetilde{F_{HT}^{VMS}} = & F_{HT}^{VMS} \underbrace{ - \frac{1}{Pe} \bigg<\pd{l^{c,h}}{x_j} \tilde{n_j} , T_D \bigg>_{\tGD}}_{\mathrm{Adjoint \; consistency \; term}} + \underbrace{ \frac{ \alpha}{h \cdot Pe} \bigg<l^{c,h} +\pd{l^{c,h}}{x_j}d_j, T_D \bigg>_{\tGD}}_{\mathrm{Penalty \; term}}, 
\end{align}

where $T_D$ is the boundary condition we want to apply, $\alpha$ is the penalty parameter for energy equation, $Pe$ is Peclet number, which is $0.7 Re$ inside our simulations, $B_{HT}^{VMS}$ is the bilinear weak form for Heat Transfer without SBM, and $F_{HT}^{VMS}$ is the linear weak form for Heat Transfer without SBM.

\subsection{Solving the CFD equations: Automated creation of meshes involving complex geometries}\label{app:mesh}
Tree-based mesh generation, using quadtrees in 2D and octrees in 3D, is common in computational sciences due to its simplicity and parallel scalability. These tree-based data structures enable efficient refinement of regions of interest, facilitating their deployment in large-scale multi-physics simulations. Our mesh generation tool, Dendro-kt~\cite{saurabh2021scalable}, provides balanced, partitioned, and parallel tree structures, making it highly effective for large-scale numerical PDE discretizations. 

\textbf{LDC (Lid-driven cavity flow):} For the LDC case, we use an octree-based mesh generation framework to create a uniform mesh with an element size of \(\frac{1}{2^9}\) over a \([0, 2] \times [0, 2]\) domain. This produces a $512 \times 512$ mesh resulting in total degress-of-freedom of $512 \times 512 \times 3 \sim 750K$. 

\begin{figure}
    \centering
    \begin{subfigure}[b]{1\linewidth}
        \includegraphics[width=\linewidth,trim=0 1.0in 0 1.0in,clip]{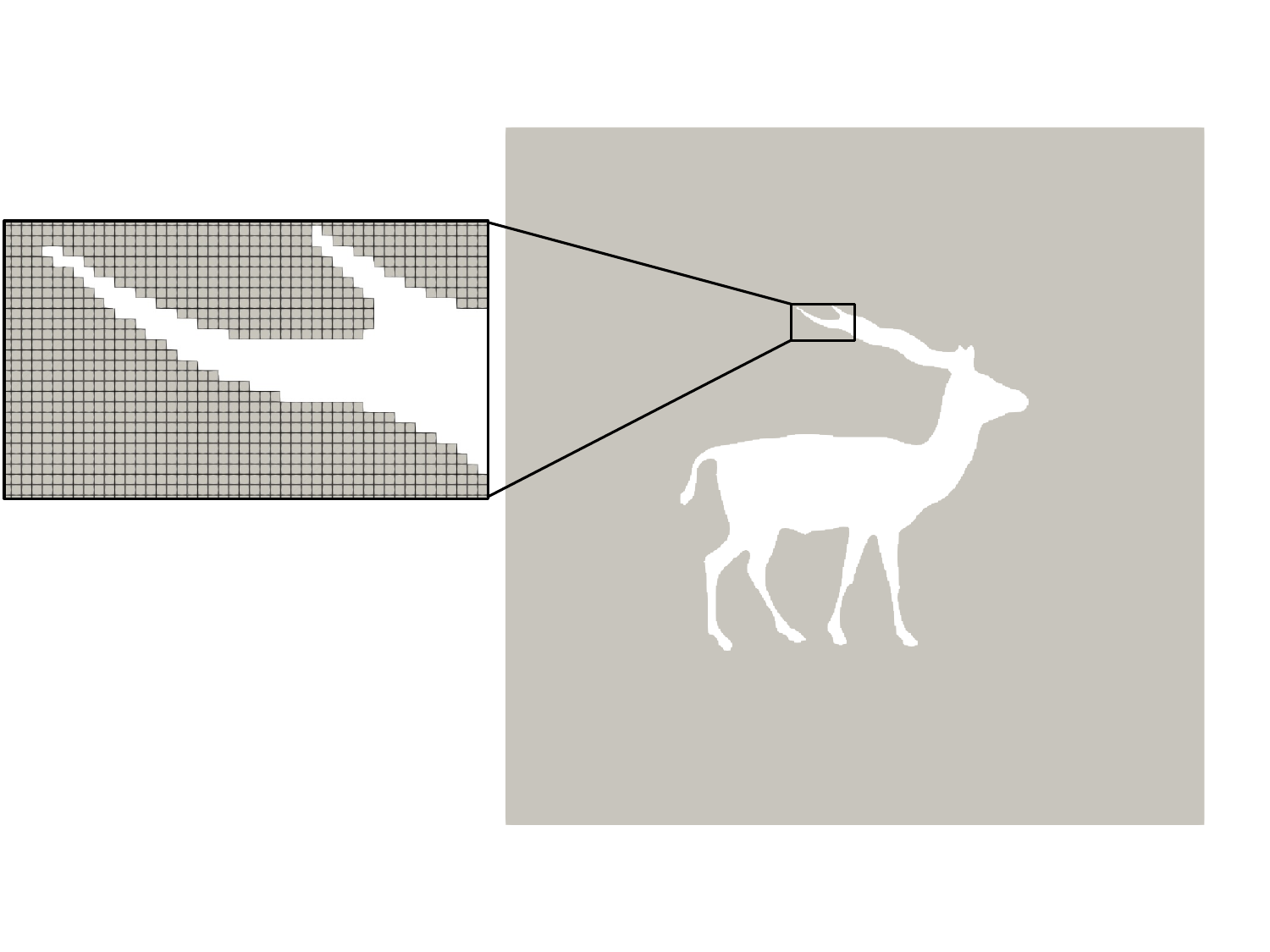}
    \end{subfigure}
    \caption{Uniform octree mesh for the lid-driven cavity flow. We illustrate the mesh with a representative shape from the Skelneton data. This mesh is carved out from a $512 \times 512$ uniform tessellation of the domain. }
    \label{fig:Mesh_LDC}
\end{figure}

\textbf{FPO (Flow passing object)}: Our computational domain is a rectangular region spanning \([0, 64] \times [0, 16]\), with the complex geometry centered at the coordinates (6, 8). Our mesh refinement strategy involves several layers of progressively coarser mesh surrounding the geometry and along the wake. Specifically, we utilize five concentric circles centered at (6, 8), with the innermost circle having a radius of 0.71 and a refinement level of 13 (yielding an element size of \(\frac{64}{2^{13}}\)), the next circle with a radius of 0.8 and a refinement level of 12 (\(\frac{64}{2^{12}}\) element size), the third circle with a radius of 1 and a refinement level of 11 (\(\frac{64}{2^{11}}\) element size), the fourth circle with a radius of 2.5 and a refinement level of 10 (\(\frac{64}{2^{10}}\) element size), and the outermost circle with a radius of 3 and a refinement level of 9 (\(\frac{64}{2^9}\) element size). We use a non-dimensional time step of 0.01 for the simulation. Starting from a non-dimensional total time of 392, we begin outputting results every 0.05 non-dimensional time units until reaching a non-dimensional total time of 404. The period from non-dimensional time 392 to 404 is when we output results that are post-processed for use in \FlowBench{}.

Additionally, we define two rectangular refinement regions aligned with the flow direction. The first rectangle has its bottom-left corner at (6, 5.5) and top-right corner at (64, 10.5), with a refinement level of 10 (\(\frac{64}{2^{10}}\) element size). The second rectangle has its bottom-left corner at (6, 5) and top-right corner at (64, 11), with a refinement level of 9 (\(\frac{64}{2^9}\) element size). Close to the geometry, to ensure detailed capture of the flow dynamics, we achieve a finer mesh with a refinement level of 14, resulting in an element size of \(\frac{64}{2^{14}}\). The mesh with different refinement levels are shown in \figref{fig:Mesh_flow_pass}.

\begin{figure}
    \centering
    \begin{subfigure}[b]{1\linewidth}
        \includegraphics[width=\linewidth,trim=0 50 0 0,clip]{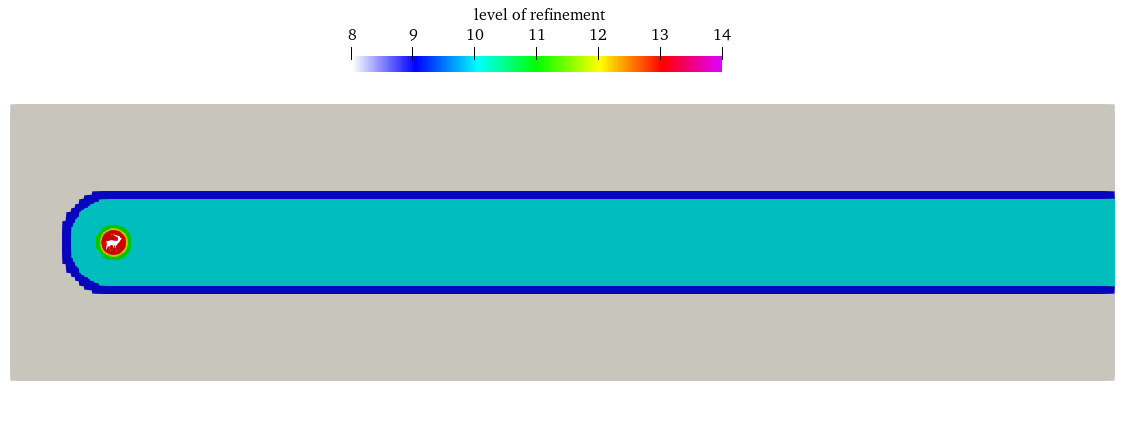}
        \caption{Overview of the mesh refinement levels.}
    \end{subfigure}
    \begin{subfigure}[b]{0.49\linewidth}
        \includegraphics[width=\linewidth,trim=0 200 0 0,clip]{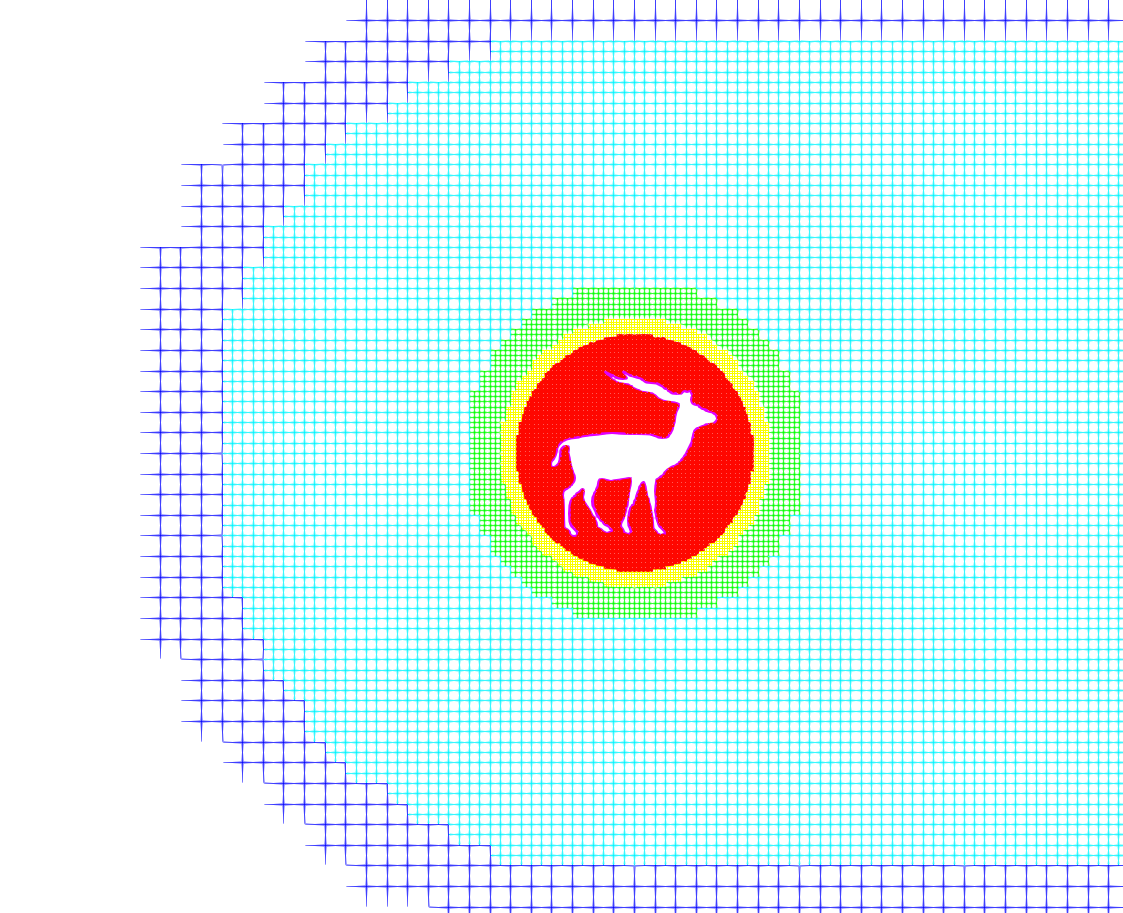}
        \caption{Zoomed-in view of the local mesh refinement. The five levels of progressively coarser meshes surrounding the geometry are clearly visible.}
    \end{subfigure}
    \hspace{0.05\linewidth}
    \begin{subfigure}[b]{0.38\linewidth}
        \includegraphics[width=\linewidth,trim=0 0 0 0,clip]{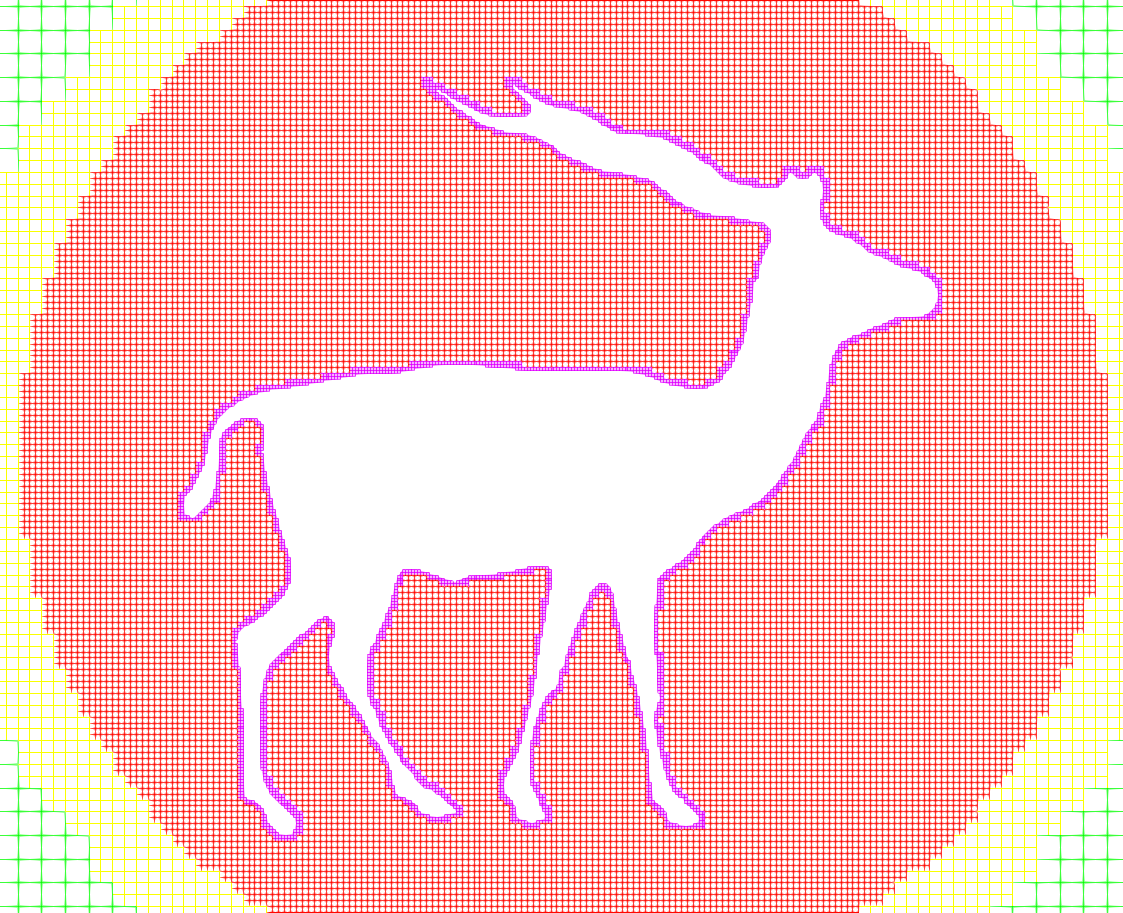}
        \caption{Zoomed-in view of the local mesh refinement near the geometry.}
    \end{subfigure}
    \caption{Local refinement octree mesh for the flow passing through the geometry. The geometry is represented by a deer.}
    \label{fig:Mesh_flow_pass}
\end{figure}

The refinement strategy leads to a total nodal point number of 117978. Consequently, the total degrees of freedom (DOFs) amount is 117978 \(\times\) 3 (353934). The combination of adaptive refinement and SBM formulation is critical to our ability to solve the PDEs in complex geometries. For comparison, if we were to use a uniform mesh (as in the case of the LDC) instead of an adaptively refined mesh, we would have had a problem with $2^{14} \times 2^{12} \times 3 \sim 201M$ DOFs!

\textbf{3D LDC:} We use a similar adaptive refinement strategy for the 3D LDC case. See \figref{fig:Mesh_3D_LDC} for an example of an object and the mesh used. Close to the object's surface, we use a refinement of level 9, producing elements of size $2/2^9$, and away from the object, we progressively coarsen the mesh to a refinement level of 7, for elements of size $2/2^7$. This produces a problem with $2.5M$ DOFs. For the boundary condition, a uniform velocity is applied along one direction to simulate the lid's movement, while the remaining two velocity components are set to zero. Specifically, the top boundary of the cavity moves with a velocity of \(u = 1\), while the transverse components \(v = 0\) and \(w = 0\), indicating no movement in those directions. All other boundaries are treated as stationary walls, where the no-slip condition applies, meaning \(u = v = w = 0\). An illustration of the boundary conditions for the 3D lid-driven cavity (LDC) is shown in \figref{fig:3D-LDC-BC}.

\begin{figure}
    \centering
    \begin{subfigure}[b]{0.45\linewidth}
        \includegraphics[width=0.7\linewidth,trim=405 0 405 0,clip]{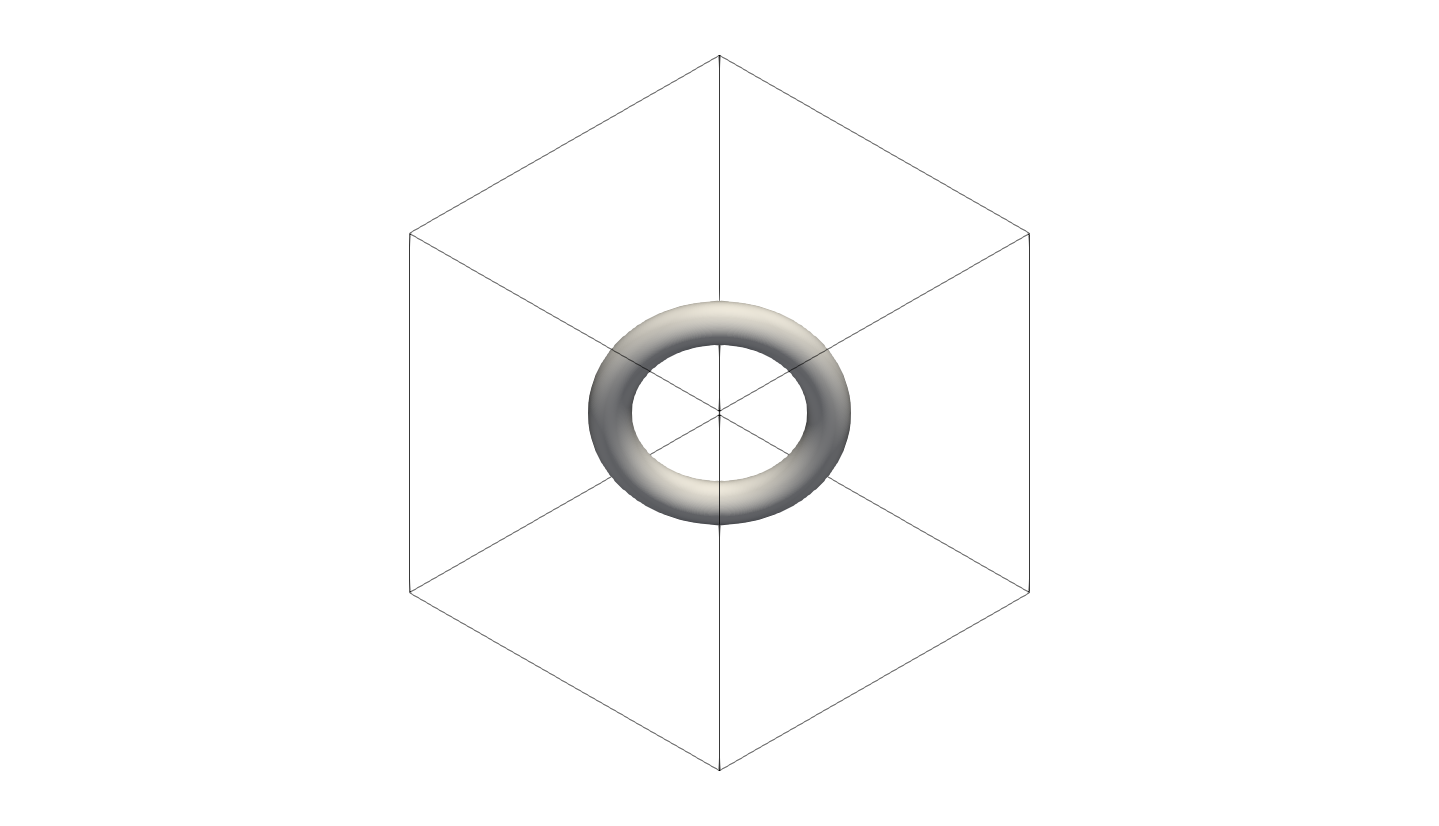}
        \caption{The object inside the lid driven cavity domain.}
    \end{subfigure}
    \hspace{0.05\linewidth}
    \begin{subfigure}[b]{0.45\linewidth}
        \includegraphics[width=1.3\linewidth,trim=0 0 0 0,clip]{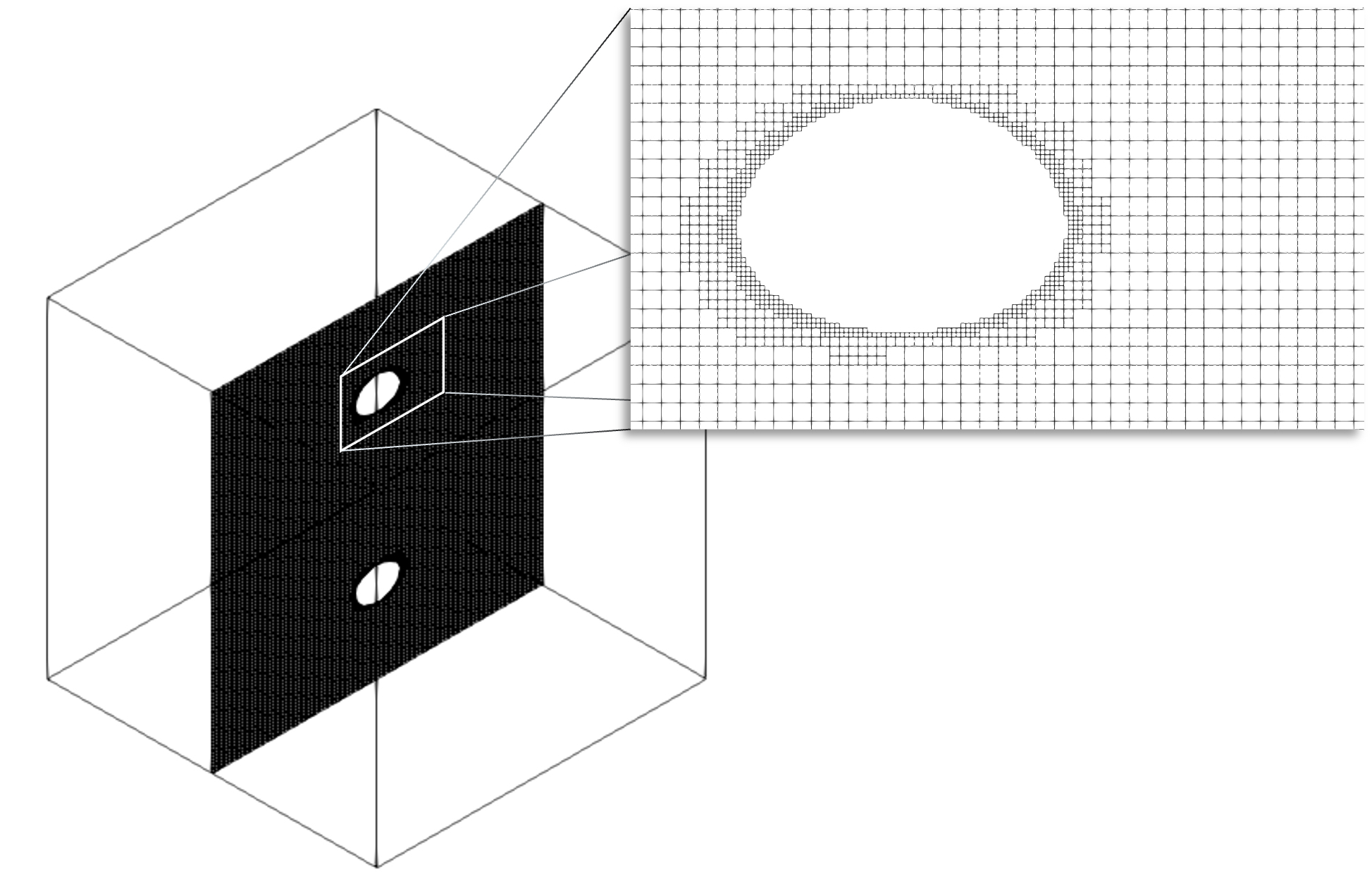}
        \caption{The octree mesh showing local refinement near the object boundary.}
    \end{subfigure}
    \caption{An example shape in the 3D LDC case, along with a slice of the computational mesh}
    \label{fig:Mesh_3D_LDC}
\end{figure}

\begin{figure}[h]
    \centering
    \includegraphics[width=0.5\textwidth]{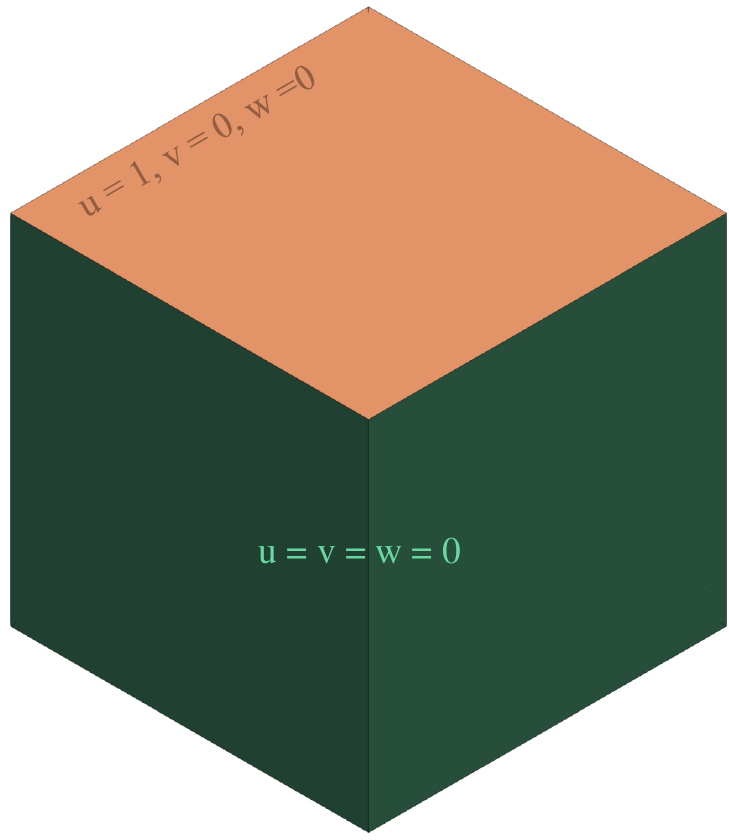}
    \caption{Boundary conditions for the 3D lid-driven cavity flow. The top boundary has \(u = 1, v = 0, w = 0\), while all other boundaries are no-slip.}
    \label{fig:3D-LDC-BC}
\end{figure}

\textbf{Solver configuration:} We use the Petsc linear algebra package for solving the system of equations. We utilize the BCGS solver, with an ASM pre-conditioner. The solver continues iterating until a relative tolerance, $rtol$, of $10^{-8}$, is reached.

\subsection{Postprocessing of the results to compute force coefficients and Nusselt number}
After the CFD solve, we also compute three engineering summary variables. The drag coefficient (\(C_D\)) represents the non-dimensional force exerted on an object in the direction of the flow. It is calculated using the formula:
\begin{align}
C_D = \frac{2 F_x}{A_{py}} .
\end{align}
where \(F_x\) is the drag force and \(A_{py}\) is the projection area in the y-direction.

The lift coefficient (\(C_L\)) indicates the non-dimensional lift force acting perpendicular to the flow direction. It is defined as:
\begin{align}
    C_L = \frac{2 F_y}{A_{px}} .
\end{align}
where \(F_y\) is the lift force and \(A_{px}\) is the projection area in the x-direction.

Finally, we compute the Nusselt number to determine the amount of heat transfer. The local Nusselt number can be defined as:
\begin{align}
Nu = \nabla T \cdot \mathbf{n} .
\end{align}

The averaged Nusselt number can be written as:
\begin{align}
\overline{Nu} = \frac{\int Nu \, d\Gamma}{\int d\Gamma} .
\end{align}
We compute the average $Nu$ across both the bottom wall as well as the surface of the object.

\subsection{Validation of the CFD framework}\label{subsec:validations}
\textbf{LDC (Lid-driven cavity flow), NS}: 
We simulate a disk with a diameter \( D = \frac{L}{3} \) placed at the center of the lid-driven cavity, where \( L \) is the chamber's box length. Our CFD results match against detailed simulations available in literature~\citep{huang2020simulation} shown in \figref{fig:VelocityProfile_LDC}.

\begin{figure}[h!]
    \centering
    \begin{subfigure}{0.49\textwidth}
    \centering
        \includegraphics[width=\linewidth,trim=0 0 0 0,clip]{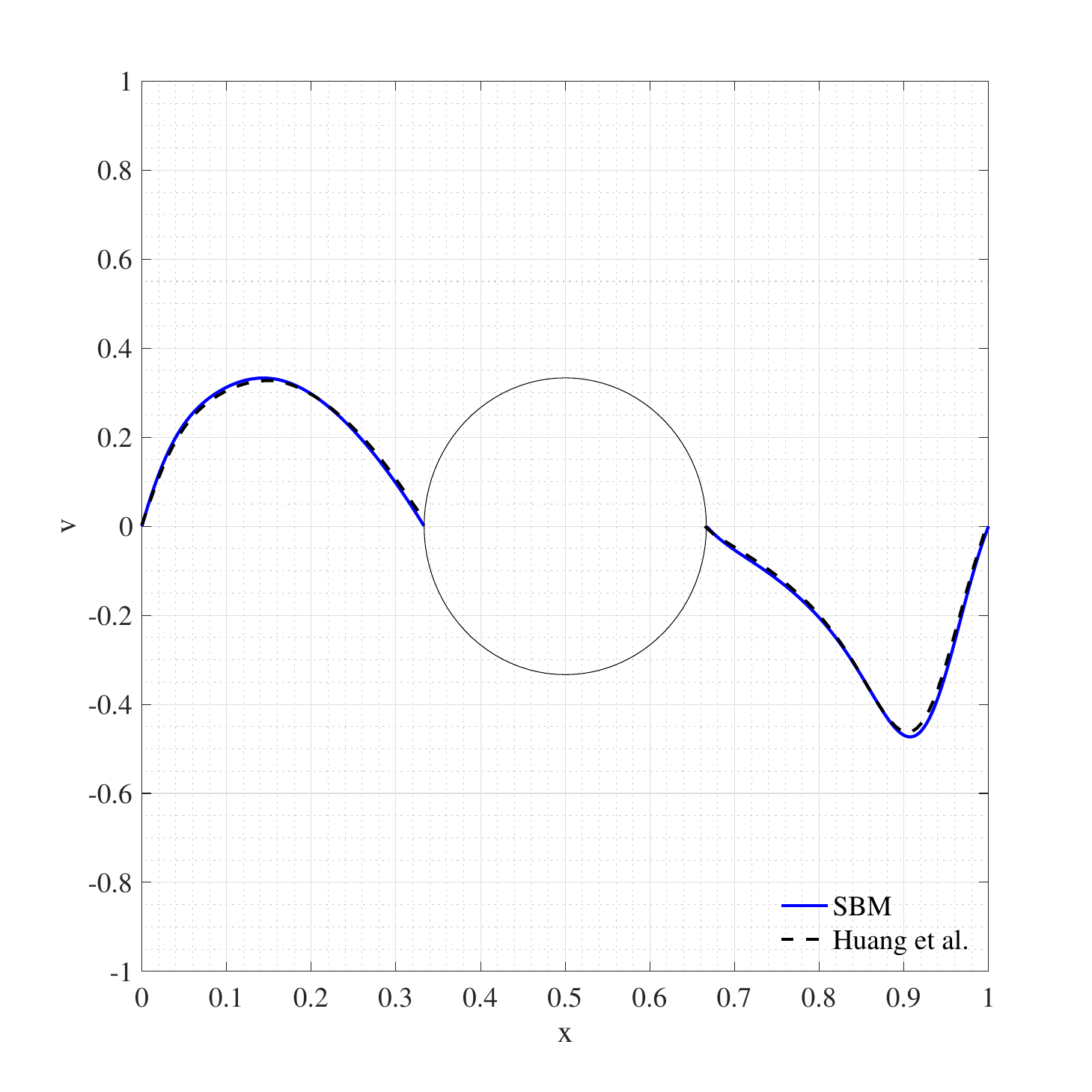}
    \caption{Horizontal velocity profile.}
    \label{fig:Quad_BoundaryFitted}
    \end{subfigure}%
    \begin{subfigure}{0.49\textwidth}
    \centering
        \includegraphics[width=\linewidth,trim=0 0 0 0,clip]{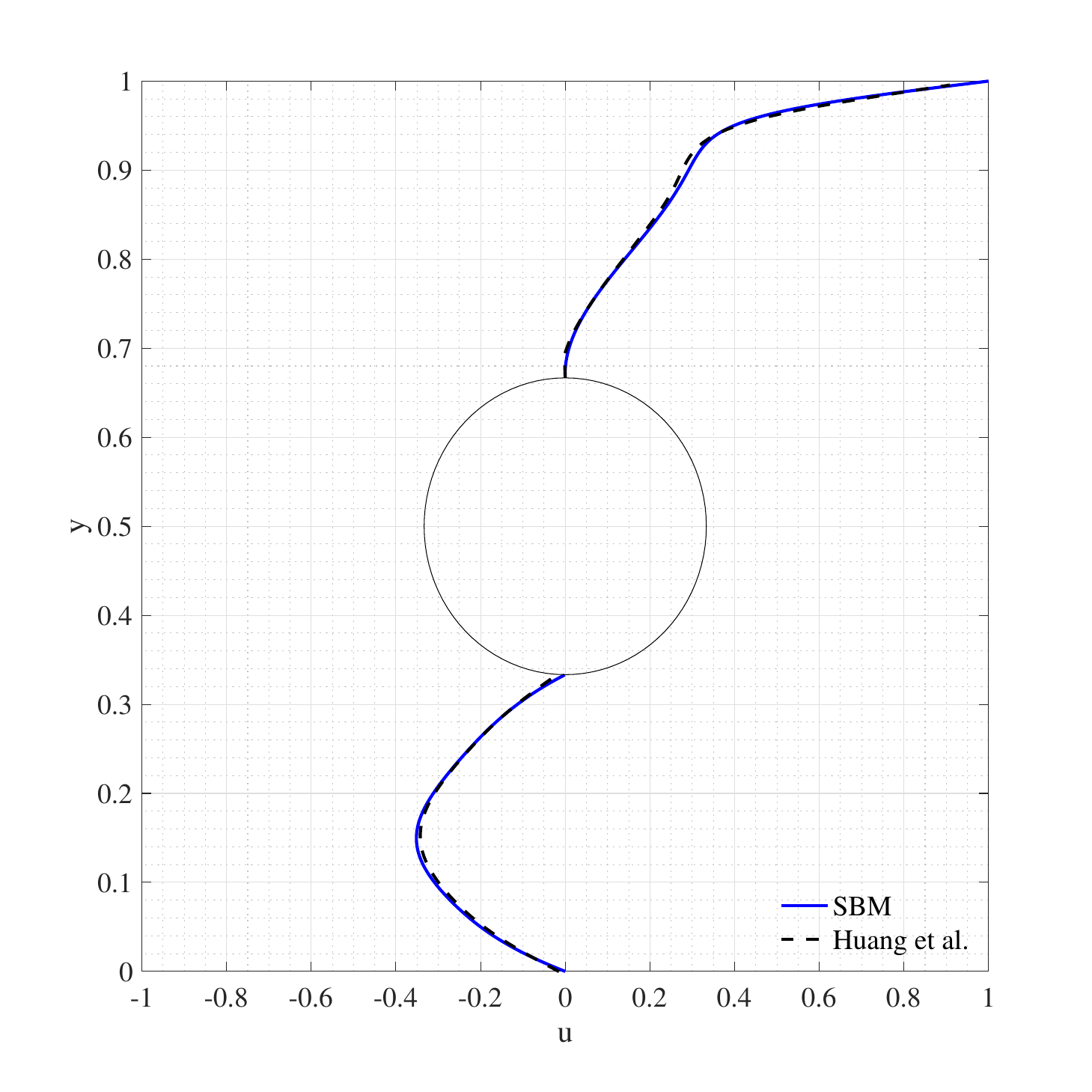}
        \caption{Vertical velocity profile.}
        \label{fig:VelocityProfile100}
    \end{subfigure}%
    \caption{Comparison of velocity profiles for lid-driven cavity flow with a circular disk at Re = 1000 against results from the literature.}
    \label{fig:VelocityProfile_LDC}
\end{figure}

\textbf{LDC (Lid-driven cavity flow), NSHT}: To validate our multiphysics simulation framework, we select a case from~\citep{Chen2020} to compare with. Here, a heated circle is placed at the center of the chamber, with a radius of 0.2L, where L represents the length of the chamber. We evaluate the local Nusselt number on the bottom wall and report an excellent match with~\citep{Chen2020} in \figref{fig:LDC_NSHT_BottomNu} at two distinct operating conditions.

\begin{figure}[h!]
\centering
\includegraphics[width=1\linewidth]{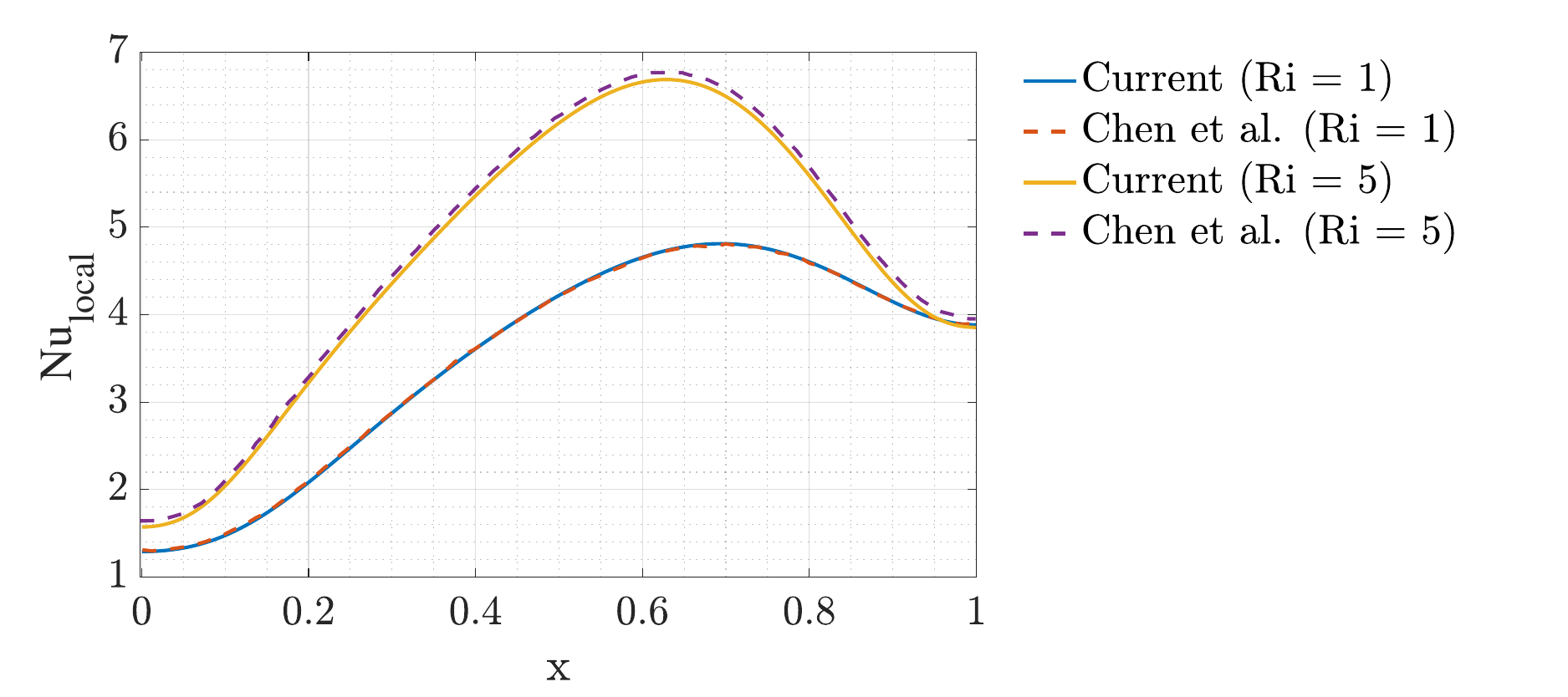}
\caption{Comparison of the Nusselt number profile along the bottom wall of the domain. The simulation results are validated against data from~\citep{Chen2020} to demonstrate the accuracy of the current multiphyics (NSHT) model.}
\label{fig:LDC_NSHT_BottomNu}
\end{figure}

\textbf{FPO (Flow passing object):} For the flow passing through geometries, we tested several shapes. To ensure the reliability of our framework, we used a circle and tested at two different Reynolds numbers: 100 and 1000. Our drag coefficients and Strouhal numbers matched well with the literature, as shown in \tabref{tab:Flow2DCircle_Re100} and \tabref{tab:Flow2DCircle_Re1000}.

\begin{table}[!t]
    \centering
    \caption{Comparison of drag coefficients and strouhal numbers for flow past a 2D cylinder at Re~=~100.}
    \begin{tabular}{@{}P{3cm}P{3cm}P{3cm}@{}}
        \toprule
                   &  \textbf{Cd} &\textbf{St}  \\      
        \midrule
         Current           &  1.35 & 0.167  \\      
         Mittal \textit{et al.}~\citep{mittal2008versatile}           &  1.35  & 0.165   \\      
         Henderson \textit{et al.}~\citep{henderson1995details}           &  1.35   & - \\      
         Luo \textit{et al.}~\citep{luo2009hybrid} & 1.35  & 0.159\\
        Kamensky \textit{et al.}~\citep{Kamensky:2015ch}      &  1.386 & 0.170  \\      
         Main \textit{et al.}~\citep{Main2018TheSB}     &  1.36 & 0.169  \\      
         Kang \textit{et al.}~\citep{kang2021variational}         &  1.374 & 0.168  \\      
        \bottomrule
    \end{tabular}
    \label{tab:Flow2DCircle_Re100}
\end{table}

\begin{table}[!t]
    \centering
    \caption{Comparison of drag coefficients and strouhal numbers
 for flow past a 2D cylinder at Re~=~1000.}
    \begin{tabular}{@{}P{3cm}P{3cm}P{3cm}@{}}
        \toprule
                   &  \textbf{Cd} &\textbf{St}  \\      
        \midrule
         Current           &  1.52 & 0.239  \\      
         Mittal \textit{et al.}~\citep{mittal2008versatile}           &  1.45  & 0.230   \\      
         Henderson \textit{et al.}~\citep{henderson1995details}           &  1.51   & - \\      
         Luo \textit{et al.}~\citep{luo2009hybrid} & 1.56  & 0.235\\
         Cheny \textit{et al.}~\citep{cheny2010ls} & 1.61  & 0.251 \\
         Jester \textit{et al.}~\citep{JESTER2003561} & 1.51 & 0.25\\
        \bottomrule
    \end{tabular}
    \label{tab:Flow2DCircle_Re1000}
\end{table}

\textbf{Complex Geometries:} To evaluate our framework's performance with complex geometries, we first present further validation of the Shifted Boundary Method (SBM) by comparing it with the Boundary-Fitted Method (BFM) in the context of lid-driven cavity simulations shown in~\figref{fig:SBM_BFM}. For the boundary-fitted mesh, we used a mesh size of approximately $\frac{1}{2^9}$, matching the element size employed in the SBM simulations.

Additionally, we validated the solver's performance for complex geometries by comparing our results to previous research. Specifically, we simulated flow past a D-shaped cylinder at various Reynolds numbers and compared the results with those from~\cite{shao2020numerical}, as shown in~\tabref{table:St}. The flow visualizations are illustrated in~\figref{fig:vis_Dshape}. Our findings closely align with the literature, demonstrating that an increase in Reynolds number corresponds to a higher Strouhal number. Readers interested in further validating our simulation framework for complex geometries are encouraged to visit our \href{https://baskargroup.bitbucket.io/#/cfdsimulation}{website}.

\begin{figure}
    \centering
    \begin{subfigure}[b]{0.45\linewidth}
        \includegraphics[width=\linewidth]{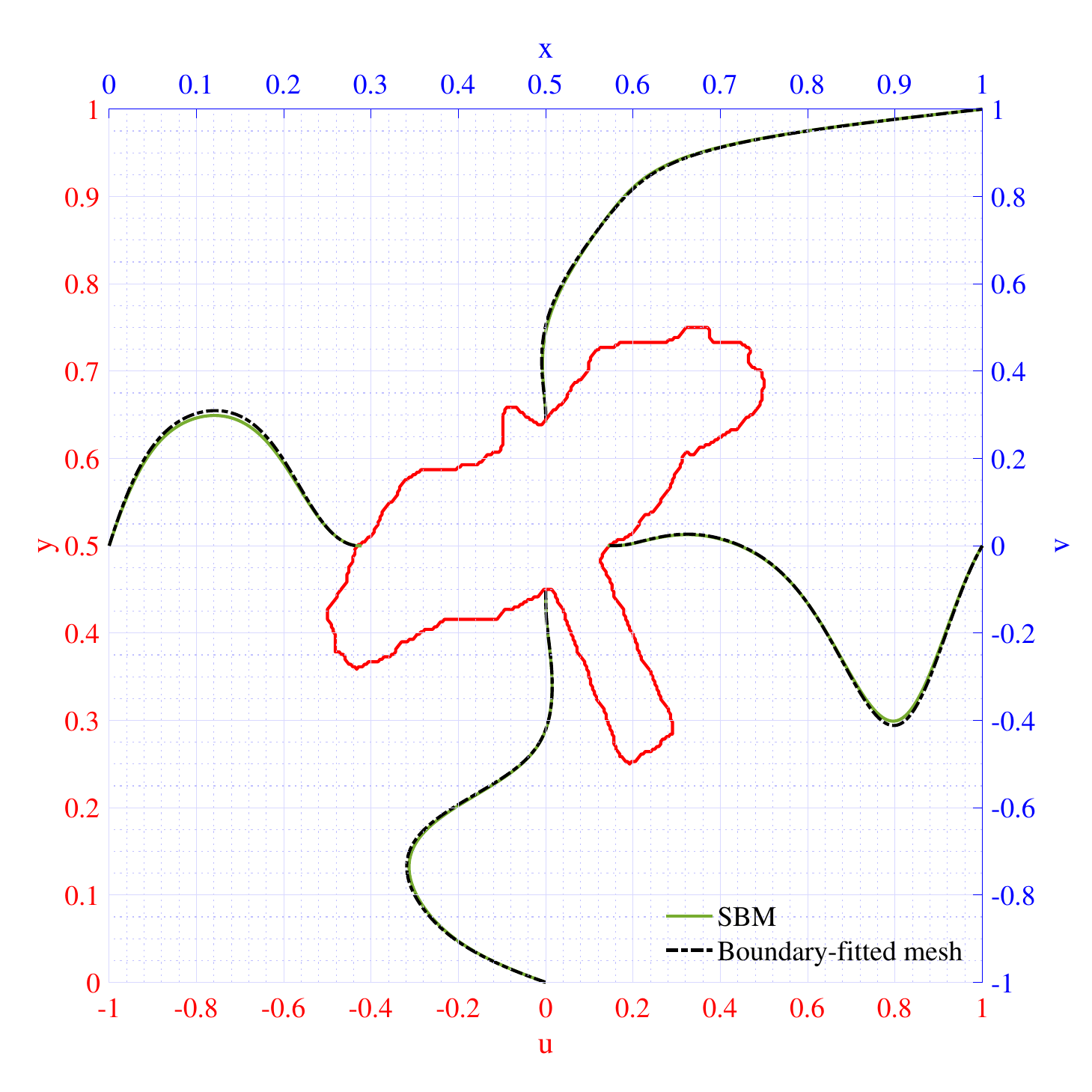}
        \caption{Comparison between SBM and BFM for the bird case at Re = 912.}
    \end{subfigure}
    \hspace{0.05\linewidth}
    \begin{subfigure}[b]{0.45\linewidth}
        \includegraphics[width=\linewidth]{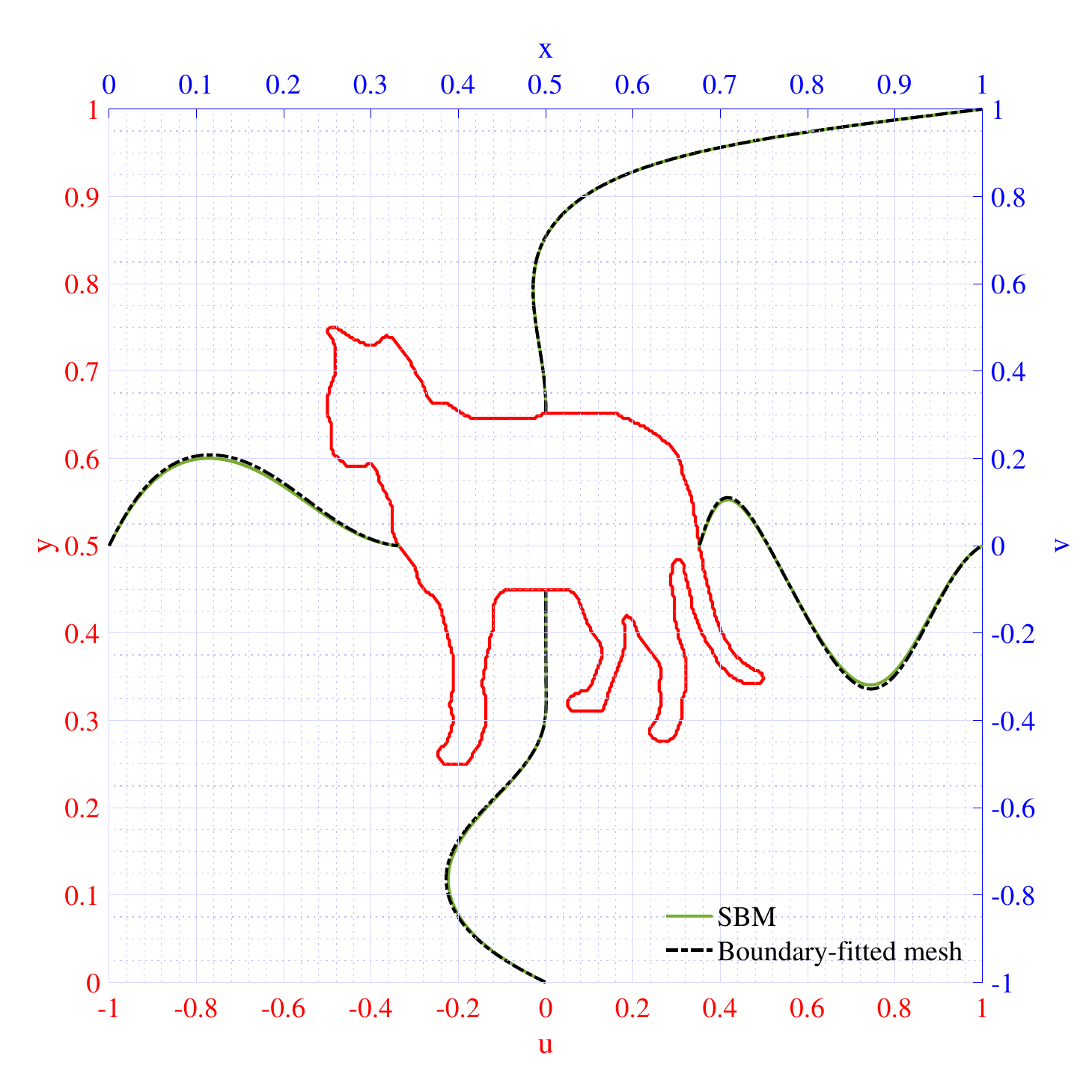}
        \caption{Comparison between SBM and BFM for the cat case at Re = 729.}
    \end{subfigure}
    \begin{subfigure}[b]{0.45\linewidth}
        \includegraphics[width=\linewidth,trim=30 60 30 60,clip]{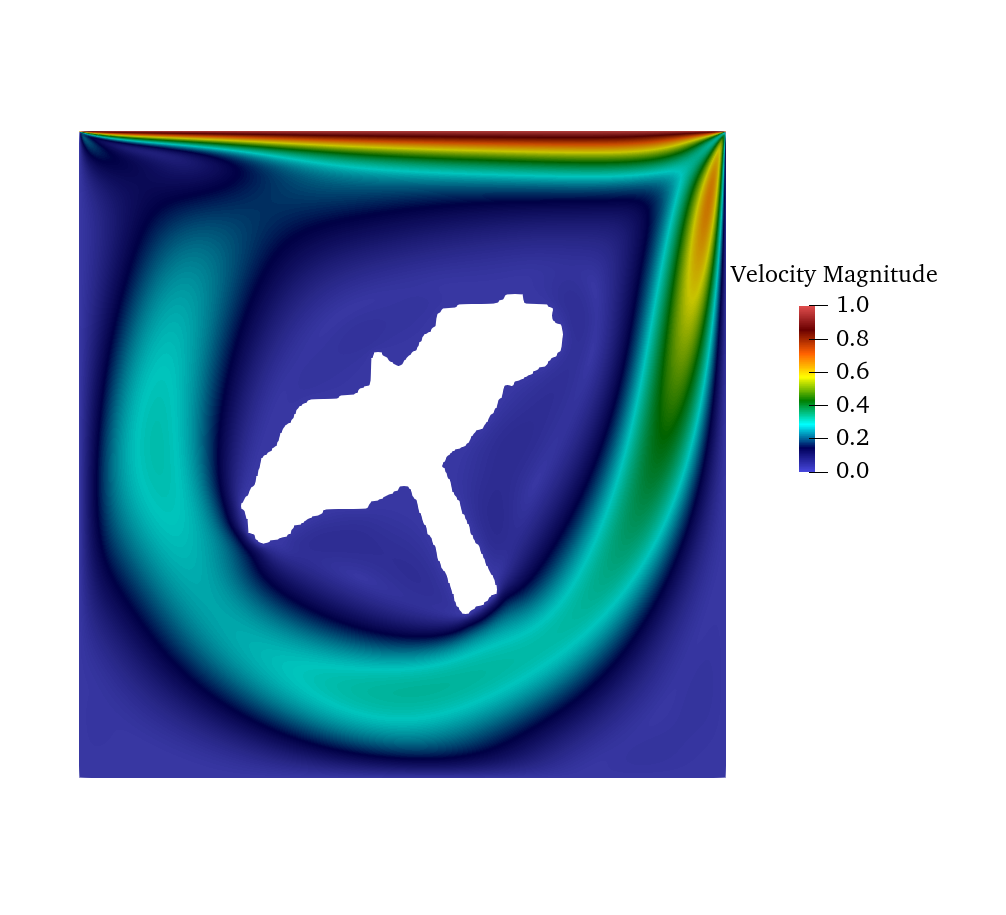}
        \caption{Flow visualization for the bird case at Re = 912.}
    \end{subfigure}
    \hspace{0.05\linewidth}
    \begin{subfigure}[b]{0.45\linewidth}
        \includegraphics[width=\linewidth,trim=30 60 30 60,clip]{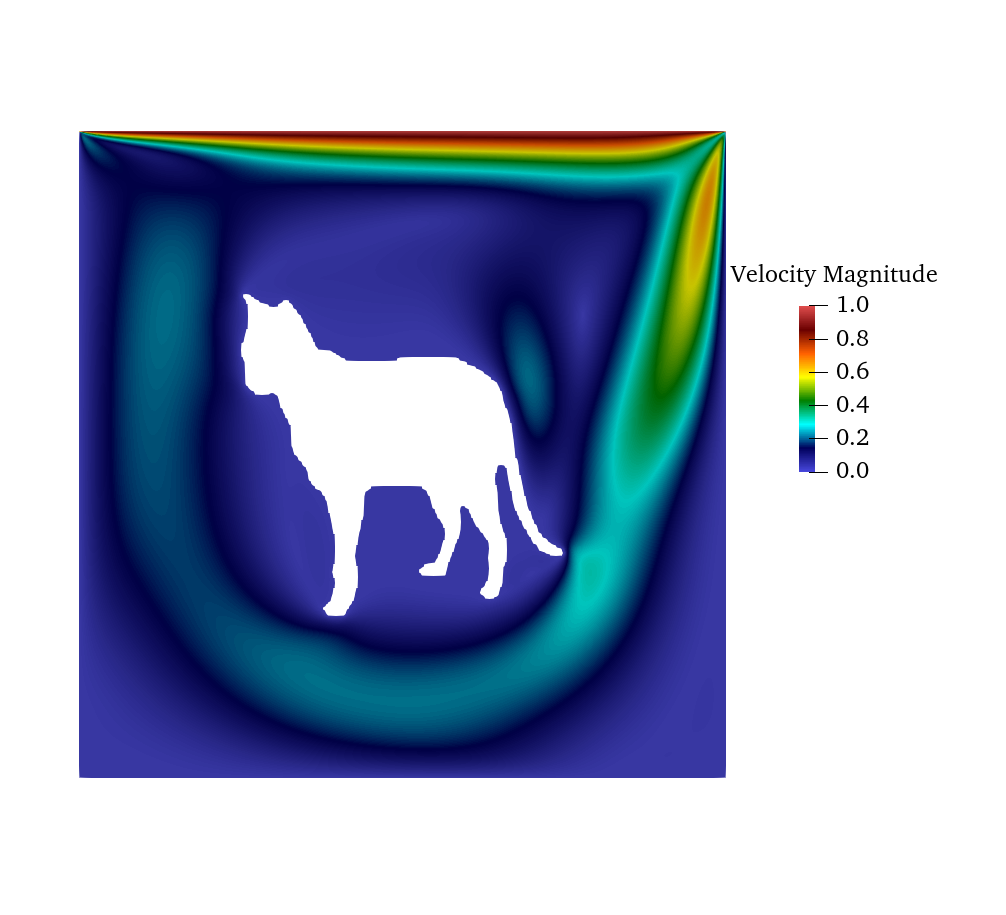}
        \caption{Flow visualization for the cat case at Re = 729.}
    \end{subfigure}
    \caption{Comparison between results from BFM and SBM in lid-driven cavity simulations for the bird and cat cases.}
    \label{fig:SBM_BFM}
\end{figure}

\begin{table}[htbp]
    \centering
    \caption{Strouhal numbers for flow past a D-shaped cylinder}
    \begin{tabular}{@{}P{1.5cm}P{1.5cm}P{3cm}@{}}
        \toprule
        \textbf{Re} & \textbf{Current} & \textbf{\cite{shao2020numerical}} \\ 
        \midrule
        46  & 0.1363 & 0.1316 \\
        100 & 0.2006 & 0.1910 \\
        160 & 0.2112 & 0.2042 \\
        \bottomrule
    \end{tabular}
    \label{table:St}
\end{table}

\begin{figure}[htbp]
    \centering
    \includegraphics[width=0.45\textwidth,trim={0 150 0 50},clip]{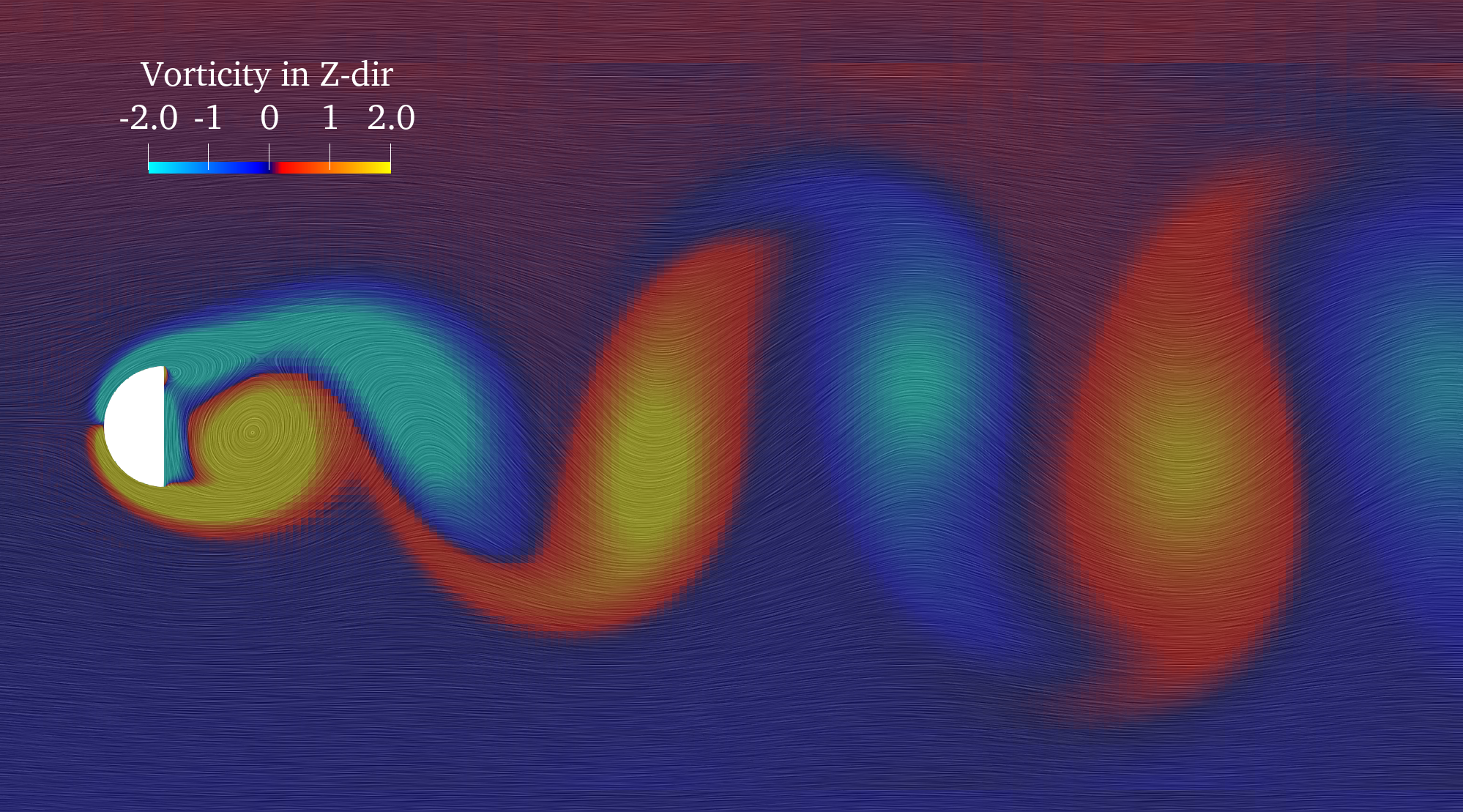}
    \caption{Flow visualization for a D-shaped cylinder at Re = 100.}
    \label{fig:vis_Dshape}
\end{figure}

\section{Benchmark Setup}
\subsection{Downsampling}
We used our in-house FEM simulation software \texttt{DendroKT} to perform forward simulations. In order to capture the physics precisely, we use a very fine resolution -- with mesh sizes at the Kolmogorov length scale in 2D, and mesh sizes at $2\times$ the Kolmogorov length scale in 3D -- to run the FEM simulations. While this approach allows us to model the physics very precisely, the resulting tensors generated at those resolutions are impractically large in size i.e., running into hundreds of gigabytes. We therefore use the ParaView tool \cite{Paraview2015} to downsample the original resolution to lower resolutions such as, $512\times512$, $256\times256$, $128\times128$, $128\times128\times128$ and $512\times128$. It is important to note that this downsampling is on the fully resolved data, and thus still captures all the larger scale features (as well as the impact of the small scale features on the large scale features).

\subsection{Machine Learning}
Our Machine Learning pipeline consists of three modules. In the first module, we undertake the tensorization exercise. We group the available downsampled data to a four or five dimensional \textit{numpy} tensor depending on the nature of the physics outlined in ~\tabref{tab:InputOutputTensors}. Tensorization helps us to subset the data using the tensor notation, and it also makes the data ready to be fed into Neural Operators/Foundation models. Data at this point is fed into the second module, where we primarily train the Neural Operators/Foundation Models. We have two principal objectives in this module, a) Learn the best hyperparameters for a given dataset and a given Neural Operator/Foundation Model. b) Select the best hyperparameter and perform a final training exercise to record the performance of each Neural Operator/Foundation Model on every dataset. Finally, we move to the post-processing module, where we report a complete suite of goodness of fit statistics using the trained model on a held-out data sample. We make our code, for select Neural Operators, publicly available.

\begin{table}[h!]
\centering
\setlength\extrarowheight{6pt}
\caption{Formulaic description of the input and output tensors. 3000/6000/1150/500 are sample sizes for the dataset. 240 is the number of equi-spaced time snapshots for the FPO case; $x,y (,z)$ are the dimensions of a field. E.g., For LDC - NS $Y[0,1,:,:]$ indicates the pointwise $v$ velocity over the entire grid for the first sample in the dataset. As stated in \secref{subsec:meta}, the FPO data requires some light postprocessing to bring it to the desired \texttt{numpy} tensor format.}
\label{tab:InputOutputTensors}
\begin{tabular}{ |c|c|c|c|} 
\hline
\textbf{Dataset} & \textbf{Dim.} &\textbf{Input Tensor} & \textbf{Output Tensor} \\
\hline
LDC - NS & 2 & $X[3000][Re, g, s][x][y]$ & $Y[3000][u,v,p][x][y]$\\
LDC - NS+HT & 2 & $X[5990][Re, Gr, g, s][x][y]$ & $Y[5990][u,v,p, \theta][x][y]$\\
FPO - NS & 2 & $X[1150][Re, g, s][x][y]$ & $Y[1150][240][u,v,p][x][y]$\\
LDC - NS & 3 & $X[500][Re, g, s][x][y][z]$ & $Y[500][u,v,p][x][y][z]$\\
\hline
\end{tabular}
\end{table}

\end{document}


\maketitle

\section{Description of Data}
\begin{table}[h!]
\centering
\setlength\extrarowheight{6pt}
\caption{Formulaic description of the input and output tensors. 3000/6000/1150/500 are sample sizes for the dataset. 240 is the number of equi-spaced time snapshots for the FPO case; $x,y (,z)$ are the dimensions of a field. E.g., $Y[0,1,:,:]$ indicates the pointwise $v$ velocity over the entire grid. We denote by $C$ - a single channel split into equal halves containing the coefficient of drag and coefficient of lift. We denote by $C^*$ - a single channel split into two equal fourths and a half containing the coefficient of drag, coefficient of lift, and the Nusselt number, respectively.}
\label{tab:InputOutputTensors}
\begin{tabular}{ |c|c|c|c|} 
\hline
\textbf{Dataset} & \textbf{Dim.} &\textbf{Input Tensor} & \textbf{Output Tensor} \\
\hline
LDC - NS & 2 & $X[3000][Re, g, s][x][y]$ & $Y[3000][u,v,p, C][x][y]$\\
LDC - NS+HT & 2 & $X[5990][Re, Gr, g, s][x][y]$ & $Y[5990][u,v,p, \theta, C^*][x][y]$\\
FPO - NS & 2 & $X[1150][Re, g, s][x][y]$ & $Y[1150][240][u,v,p][x][y]$\\
LDC - NS & 3 & $X[500][Re, g, s][x][y][z]$ & $Y[500][u,v,w,p][x][y][z]$\\
\hline
\end{tabular}
\end{table}

\section{Datasheet for our Dataset}
We follow the datasheet proposed in \cite{Gebru2021} for documenting our \FlowBench{} dataset. 


\begin{enumerate}
    \item Motivation 
    \begin{enumerate}[label=(\alph*)]
        \item For what purpose was the dataset created? \\
        The dataset was created to enable comprehensive evaluation of data-driven models that predict complex 2D/3D flow physics around a range of geometrical objects in steady-state and transient situations.
        
        \item Who created the dataset and on behalf of which entity?\\
        The dataset was created by the Baskar Ganapathysubramanian Group (ComPM Lab) at Iowa State University, Ames, IA.
        \item Who funded the creation of the dataset? \FlowBench was funded in part by the the AI Research Institutes program supported by USDA-NIFA under AI Institute: for Resilient Agriculture, Award No. 2021-67021-35329, NSF under awards 1954556, 2323716, and 2053760.
        \item Any other Comments?\\
        \textbf{\answerNA{}}
    \end{enumerate}

    \item Composition
    \begin{enumerate}[label=(\alph*)]
        \item What do the instances that comprise the dataset represent?\\
        Each instance captures some cardinal flow variables, such as velocities, pressure, or temperature defined over the entire domain of interest. In addition, each such field is packaged together as a single tensor, the details of which are discussed in Details in \tabref{tab:InputOutputTensors}.
        \item How many instances are there in total?\\
        10,650
        \item Does the dataset contain all possible instances or is it a sample (not necessarily random) of instances from a larger set?\\
        We sample across geometries and flow operating conditions. Geometries are further classified into parametric and non-parametric families. Flow operating conditions are characterized by non-dimensional numbers (Reynolds and Grashof). Our dataset includes a dense, random collection of flow simulations sampled from this space of geometry $\times$ operating conditions. However, given that the space is a continuous space, it cannot contain all possible instances. For this reason, we have provided the complete code to the end user to generate more random geometries and, by implication, more input data instances. 
        \item What data does each instance consist of?\\
        Each data is a simulation outcome. \\ The simulation inputs are the shape of the object, and the operating conditions. The object's shape is provided in two (field) formats: binary mask and signed distance field (SDF). The operating conditions -- Reynolds number, $Re$, and Grashof number $Gr$ -- are provided as concordant fields. If $Gr$ is not provided, that simulation was performed for $Gr=0$. \\ The simulation output consists of (a) fluid velocity in the domain given as a 2- or 3- dimensional field ($u, v, w)$, (b) pressure in the domain given as a field, (c) temperature (if flow thermal simulation case) in the domain given as a field, (d) Coefficient of lift, coefficient of drag, and Nusselt number as concordant fields 
        
        In \tabref{tab:InputOutputTensors}, we have presented a detailed tensor formula that completely explains the constitution of each sample instance present in our dataset.
        \item Is there a label or target associated with each instance?\\
        See the previous point. 
        \item Is any information missing from individual instances?\\
        \textbf{\answerNo{}}
        \item Are relationships between individual instances made explicit?\\
        \textbf{\answerYes{}}. Details in \tabref{tab:InputOutputTensors}.
        \item Are there recommended data splits?\\
        \textbf{\answerNo{}}
        \item Are there any errors, sources of noise, or redundancies in the dataset?\\
        \textbf{\answerNo{}}
        \item Is the dataset self-contained, or does it link to or otherwise rely on external resources (e.g., websites, tweets, other datasets)?\\
        \textbf{\answerYes{}}. Our dataset is self-contained.
        \item Does the dataset contain data that might be considered confidential (e.g., data that is protected by legal privilege or by doctor-patient confidentiality, data that includes the content of individuals’ non-public communications)?\\
        \textbf{\answerNo{}}
        \item Does the dataset contain data that, if viewed directly, might be offensive, insulting, threatening, or might otherwise cause anxiety?\\
        \textbf{\answerNo{}}
        \item Does the dataset relate to people?\\
        \textbf{\answerNo{}}
        \item Does the dataset identify any subpopulations (e.g., by age, gender)?\\
        \textbf{\answerNo{}}
        \item Is it possible to identify individuals (i.e., one or more natural persons), either directly or indirectly (i.e., in combination with other data) from the dataset?\\
        \textbf{\answerNo{}}
        \item Does the dataset contain data that might be considered sensitive in any way (e.g., data that reveals racial or ethnic origins, sexual orientations, religious beliefs, political opinions or union memberships, or locations; financial or health data; biometric or genetic data; forms of government identification, such as social security numbers; criminal history)?\\
        \textbf{\answerNo{}}
       \item Any other Comments?\\
       \textbf{\answerNo{}}
    \end{enumerate}

    \item Collection Process
    \begin{enumerate}[label=(\alph*)]
        
        \item How was the data associated with each instance acquired?\\
        Given a set of flow conditions and geometries, detailed flow  simulations (using an in house validated simulator) were run and monitored to ensure results were properly convergent.
        
        \item What mechanisms or procedures were used to collect the data (e.g., hardware apparatus or sensor, manual human curation, software program, software API)?\\
        Flow simulations -- specifically finite element simulation -- were used to generate the data using our in-house simulation framework~\cite{saurabh2021scalable}.
        
        \item If the dataset is a sample from a larger set, what was the sampling strategy (e.g., deterministic, probabilistic with specific sampling probabilities)?\\
        See Point 2(c). The flow conditions are randomly generated, using a Sobel sequence generator.
        \item Who was involved in the data collection process (e.g., students, crowdworkers, contractors), and how were they compensated (e.g., how much were crowdworkers paid)?\\
        Only the authors of the paper were involved. No third parties or contract workers were involved.
        \item Over what timeframe was the data collected?\\
        May 2024 to June 2024.
        \item Were any ethical review processes conducted (e.g., by an institutional review board)?\\
        \textbf{\answerNA{}}
        \item Does the dataset relate to people?\\
        \textbf{\answerNo{}}
        \item Did you collect the data from the individuals in question directly, or obtain it via third parties or other sources (e.g., websites)?\\
        \textbf{\answerNA{}}
        \item Were the individuals in question notified about the data collection?\\
        \textbf{\answerNA{}}
        \item Did the individuals in question consent to the collection and use of their data?\\
        \textbf{\answerNA{}}
        \item If consent was obtained, were the consenting individuals provided with a mechanism to revoke their consent in the future or for certain uses?\\
        \textbf{\answerNA{}}
        \item Has an analysis of the potential impact of the dataset and its use on data subjects (e.g., a data protection impact analysis) been conducted?\\
        \textbf{\answerNA{}}
        \item Any other Comments?\\
        \textbf{\answerNo{}}
    \end{enumerate}

    \item \textbf{Preprocessing, Cleaning and Labeling}
    \begin{enumerate}[label=(\alph*)]
        \item Was any preprocessing/cleaning/labeling of the data done (e.g., discretization or bucketing, tokenization, part-of-speech tagging, SIFT feature extraction, removal of instances, processing of missing values)?\\
        All our raw data was generated using our in-house FEM software. The raw data was (sub)sampled and cropped using the ParaView tool \cite{Paraview2015}. These processed fields were packaged as numpy tensors for immediate use in Deep Learning frameworks.
        \item Was the "raw" data saved in addition to the preprocessed/cleaned/labeled data (e.g., to support unanticipated future uses)?
        \textbf{\answerYes{}}
        \item Is the software used to preprocess/clean/label the instances available? \\
        We used either the ParaView tool \cite{Paraview2015} or in-house scripts - released in the "Code" section of our \href{https://baskargroup.bitbucket.io/}{website}
        \item Any other Comments?\\
        \textbf{\answerNo{}}
    \end{enumerate}

    \item Uses
    \begin{enumerate}[label=(\alph*)]
        \item Has the dataset been used for any tasks already?\\
        \textbf{\answerYes{}}. We have used the dataset to benchmark three leading Neural Operator models in the manuscripts.
        \item Is there a repository that links to any or all papers or systems that use the dataset?\\
        \textbf{\answerNo{}}
        \item What (other) tasks could the dataset be used for?\\
        \textbf{\answerNA{}}
        \item Is there anything about the composition of the dataset or the way it was collected and preprocessed/cleaned/labeled that might impact future uses?\\
        \textbf{\answerNo{}}
        \item Are there tasks for which the dataset should not be used?\\
        \textbf{\answerNo{}}
        \item Any other Comments?\\
        \textbf{\answerNo{}}
    \end{enumerate}

    \item Distribution
    \begin{enumerate}[label=(\alph*)]
        \item Will the dataset be distributed to third parties outside of the entity (e.g., company, institution, organization) on behalf of which the dataset was created?\\
        \textbf{\answerYes{}}
        \item How will the dataset be distributed (e.g., tarball on website, API, GitHub)?\\
        We use an institutional data distributional service provided by Iowa State University, Ames, IA. This service allows direct download of our datasets using a URL, which has been included in our main paper.
        \item When will the dataset be distributed?\\
        Upon acceptance of the paper.
        \item Will the dataset be distributed under a copyright or other intellectual property (IP) license, and/or under applicable terms of use (ToU)?\\
        Licensed under CC BY-NC 4.
        \item Have any third parties imposed IP-based or other restrictions on the data associated with the instances?\\
        \textbf{\answerNA{}}. Third parties are not involved.
        \item Do any export controls or other regulatory restrictions apply to the dataset or to individual instances?\\
        \textbf{\answerNA{}}
        \item Any other Comments?\\
        \textbf{\answerNo{}}
    \end{enumerate}

    \item Maintenance
    \begin{enumerate}[label=(\alph*)]
        \item Who is supporting/hosting/maintaining the dataset?\\
        Baskar Group (ComPM Lab) at Iowa State University, Ames, IA.
        \item How can the owner/curator/manager of the dataset be contacted (e.g., email address)?\\
        The corresponding author of the main paper will be the single point of contact. 
        \item Is there an erratum?\\
        \textbf{\answerNo{}}
        \item Will the dataset be updated (e.g., to correct labeling errors, add new instances, delete instances’)?\\
        \textbf{\answerNo{}}
        \item If the dataset relates to people, are there applicable limits on the retention of the data associated with the instances (e.g., were individuals in question told that their data would be retained for a fixed period of time and then deleted)?\\
        \textbf{\answerNA{}}
        \item Will older versions of the dataset continue to be supported/hosted/maintained?\\
        \textbf{\answerNo{}}. Since this is a physics dataset, we do not anticipate any changes to the content of our dataset.
        \item If others want to extend/augment/build on/contribute to the dataset, is there a mechanism for them to do so?\\
        \textbf{\answerNo{}}
        \item Any other Comments?\\
        \textbf{\answerNo{}}
    \end{enumerate}
\end{enumerate}

\section{URLs for Dataset Access}
\begin{itemize}
    \item \textbf{Project Website:} \url{https://baskargroup.bitbucket.io}
    \item \textbf{Hosting Dataset:} \url{https://figshare.com/s/15e9d23790d0a14e8f71}
    \item \textbf{Code:} \url{https://github.com/baskargroup/flowbench-tools}
\end{itemize}

\section{Author Responsibility Statement}

As the authors of this submission, we affirm that we bear all responsibility in case of any rights violations or ethical issues associated with this work. We confirm that the submitted work is original, and if it includes third-party content (e.g., text, data, software), it is used with proper permissions and attributions. The authors assume full responsibility in case of violation of any rights. \\

\subsection{Data License Confirmation}

We confirm that all data used in this research has been generated by us. Furthermore, we confirm that the use of data in this research adheres to all applicable laws and ethical guidelines. In addition, we make our data available for general use under the terms of the CC BY-NC 4.0 license\\

\section{Hosting, Licensing and Maintenance Plan}
The authors are hosting all the data at the data release portal, with the URL available in the main paper. The data release portal is hosted by Iowa State University, Ames, IA, under an institutional license from Figshare. All the data made available will continue to be maintained by the authors at the aforesaid portal under the DOI - 10.25380/iastate.25939561

\bibliographystyle{unsrtnat}
\bibliography{references}